\documentclass[a4paper,12pt]{article}
\usepackage{amsmath,amssymb}
\usepackage{bbm}
\usepackage[nosort]{cite}
\usepackage{graphicx}
\usepackage[bookmarks=true]{hyperref}
\usepackage{feynmp}

\usepackage{ifpdf}
\ifpdf
\fi

\ifx\hypersetup\sadfkjashdfkxja\else
\hypersetup{pdftitle={}}
\hypersetup{pdfauthor={}}
\hypersetup{pdfsubject={}}
\hypersetup{pdfkeywords={}}
\hypersetup{plainpages=false}
\hypersetup{bookmarksnumbered=true}
\hypersetup{pdfstartview=FitH}
\hypersetup{pdfpagemode=UseNone}
\hypersetup{colorlinks=false}
\hypersetup{citebordercolor={.5 1 .5}}
\hypersetup{urlbordercolor={.5 1 1}}
\hypersetup{linkbordercolor={1 .7 .7}}
\fi

\usepackage[a4paper,left=2.8cm,right=2.8cm,top=2.8cm,bottom=2.8cm]{geometry}

\newcommand{\matr}[2]{\left(\begin{array}{#1}#2\end{array}\right)}

\newcommand{\tprods}[2]{\langle#1#2\rangle}

\newcommand{\Tr}{\mathop{\mathrm{Tr}}}

\def\ni{\noindent}
\def\be{\begin{equation}}
\def\ee{\end{equation}}
\def\bsp{\be\begin{split}}
\def\la{\langle}
\def\ra{\rangle}
\def\dag{\dagger}

\def\lr{\leftrightarrow}

\def\G{\Gamma}
\def\D{\Delta}

\def\a{\alpha}
\def\b{\beta}
\def\g{\gamma}

\def\d{\delta}
\def\e{\epsilon}
\def\m{\mu}
\def\tr{\mbox{tr}}
\def\n{\nu}
\def\s{\sigma}
\def\r{\rho}

\def\t{\tau}

\def\p{\partial}

\def\lr{\leftrightarrow}

\def\w{\wedge}

\begin{document}
\begin{fmffile}{3D1}

\newpage
\setcounter{page}{1}
\pagenumbering{arabic}
\renewcommand{\thefootnote}{\arabic{footnote}}
\setcounter{footnote}{0}

\title{\vspace{-2cm}
       \hfill
       \mbox{\normalsize{NORDITA-2013-15}}
       \vskip 20pt
Scattering Amplitudes of Massive $\mathcal{N} = 2$ Gauge
  Theories in Three Dimensions}

\author{Abhishek
  Agarwal\footnote{Email: abhishek@aps.org}$^{~1}$, Arthur
  E. Lipstein\footnote{Email: Arthur.Lipstein@maths.ox.ac.uk}$^{~2}$,
  Donovan Young\footnote{Email: donovany@kth.se}$^{~3}$ \\ \\ $^1$
  Physical Review Letters, American Physical Society, \\1 Research
  Road, Ridge, NY 11961, USA\\ and\\ Physics Department, City College
  of CUNY, New York, NY 10031 USA \\ \\ $^2$ The Mathematical
  Institute, University of Oxford,\\ 29-29 St Giles', Oxford, OX1 3LB,
  UK \\ \\ $^3$ Nordita, KTH Royal Institute of Technology and
  Stockholm University,\\ Roslagstullsbacken 23, SE-106 91 Stockholm,
  Sweden \\ }

\maketitle
\begin{abstract}
We study the scattering amplitudes of mass-deformed Chern-Simons
theories and Yang-Mills-Chern-Simons theories with $\mathcal{N}=2$
supersymmetry in three dimensions. In particular, we derive the
on-shell supersymmetry algebras which underlie the scattering matrices
of these theories. We then compute various 3 and 4-point on-shell
tree-level amplitudes in these theories. For the mass-deformed
Chern-Simons theory, odd-point amplitudes vanish and we find that all
of the 4-point amplitudes can be encoded elegantly in 
superamplitudes. For the Yang-Mills-Chern-Simons theory, we obtain all
of the 4-point tree-level amplitudes using a combination of
perturbative techniques and algebraic constraints and we comment on
difficulties related to computing amplitudes with external gauge
fields using Feynman diagrams.  Finally, we propose a BCFW recursion
relation for mass-deformed theories in three dimensions and discuss
the applicability of this proposal to mass-deformed $\mathcal{N}=2$
theories.
    
{\normalsize \par}
\end{abstract}

\newpage
\tableofcontents

\section{Introduction}
Over the past few years, there has been a great deal of progress in
understanding the scattering amplitudes of three-dimensional gauge theories. The
study of scattering amplitudes of Chern-Simons-Matter theories
with $\mathcal{N} \geq 4$ supersymmetry and Supersymmetric Yang-Mills
theories (SYM) with $\mathcal{N} \geq 2$ supersymmetry (initiated in
\cite{SCS} and \cite{Chiou:2005jn,AADY1,Lipstein:2012kd} respectively)
shows that the $S$-matrices of three dimensional supersymmetric gauge
theories contain fascinating simplifying aspects that are not manifest
in their traditional Lagrangian descriptions. For instance, it was
shown in \cite{SCS} that the four-particle amplitudes of a large
family of Chern-Simons-Matter theories have the same formal structure as the
scattering matrix of the spin chain that is the large-$N$ dilatation
operator of $\mathcal{N} = 4$ SYM in $d=4$. Furthermore, amplitudes of
the $\mathcal{N}=8$ superconformal Chern-Simons theory known as the
BLG theory \cite{Bagger:2007jr,Gustavsson:2007vu} were studied in
\cite{Huang:2010rn}.

More recently, a BCFW recursion relation
\cite{Britto:2005fq,ArkaniHamed:2010kv} for three-dimensional gauge
theories with massless fields was developed in \cite{Gang:2010gy}, and
used to show that an $\mathcal{N}=6$ superconformal Chern-Simons
theory known as the ABJM theory \cite{Aharony:2008ug} has dual
superconformal symmetry both at tree \cite{Bargheer:2010hn,
  Huang:2010qy} and loop-level. Dual superconformal symmetry
\cite{Drummond:2006rz,Brandhuber:2008pf,Drummond:2008vq} is
inequivalent to ordinary superconformal symmetry and generates Yangian
symmetry when combined with ordinary superconformal symmetry
\cite{Dolan:2004ps}. In 4d $\mathcal{N}=4$ SYM, dual superconformal
symmetry corresponds to the ordinary superconformal symmetry of
null-polygonal Wilson loops that are dual to the amplitudes
\cite{Alday:2007hr,Drummond:2007cf,Drummond:2007aua,Mason:2010yk,CaronHuot:2010ek}. The
Yangian symmetry of $\mathcal{N}=4$ SYM can be made manifest using a
Grassmannian integral formula developed in
\cite{ArkaniHamed:2009dn}. An analogous formula for the ABJM theory
was proposed in \cite{Lee:2010du}. This formula involves an integral
over the orthogonal Grassmannian. Some evidence for an amplitude/Wilson
loop duality in the ABJM theory was found in
\cite{Henn:2010ps,Chen:2011vv,Bianchi:2011dg}. Recently, 1-loop
amplitudes were computed in the ABJM theory and shown to exhibit new
structures which do not appear in 4d $\mathcal{N}=4$ SYM theory,
notably sign functions of the kinematic variables
\cite{Bianchi:2012cq,Bargheer:2012cp,Brandhuber:2012un}.

The recursion relation proposed in \cite{Gang:2010gy} was also used to
show that three-dimensional maximal SYM amplitudes have dual conformal
covariance \cite{Lipstein:2012kd}. Note that three-dimensional SYM
theories do not have ordinary superconformal symmetry because the
Yang-Mills coupling constant is dimensionful in three
dimensions. Three-dimensional SYM theories exhibit a number of other
surprising properties. In particular, references \cite{AADY1,Lipstein:2012kd}
showed that they have helicity structure and reference \cite{AADY1}
showed that their 4-point amplitudes have enhanced $R$-symmetry which
originates from the duality between scalars and vectors in three
dimensions. Furthermore, the loop amplitudes of three-dimensional
maximal SYM theory have a similar structure to those of the ABJM
theory. In particular, 1-loop corrections are finite or vanish in both
theories \cite{Lipstein:2012kd}. Furthermore, the 2-loop 4-point
amplitudes of both theories can be matched in the Regge-limit
\cite{Bianchi:2012ez}. It was recently shown that three-dimensional
supergravity amplitudes can be obtained as double copies of both
three-dimensional supersymmetric Chern-Simons theories and
three-dimensional SYM theories
\cite{Bargheer:2012gv,Huang:2012wr}. All these remarkable developments
provide ample motivation for further investigation into the
scattering amplitudes of gauge theories in three spacetime dimensions.

It may be fair to say that most of the investigations mentioned above
are largely confined to the studies of massless theories with high
degrees of supersymmetry. In this paper, we explore a complementary
part of the landscape of $d=3$ gauge theories from the point of view
of scattering amplitudes. Specifically, we investigate mass-deformed
$\mathcal{N}=2$ gauge theories with adjoint matter fields. The two
theories that span this category are mass-deformed Chern-Simons theory
(hereafter referred to as CSM theory) and Yang-Mills-Chern-Simons
theory with $\mathcal{N}=2$ supersymmetry (YMCS), and we investigate their
tree-level color-ordered scattering amplitudes in this paper. Whereas
the gauge field has no propagating degrees of freedom in the
Chern-Simons theory, in the Yang-Mills-Chern-Simons theory the gauge
field has one propagating degree of freedom, which is massive. In
particular, the Chern-Simons term provides a topological mass for the
gauge field without breaking gauge invariance or locality
\cite{Deser:1981wh,Deser}.

From the point of view of scattering amplitudes, these theories are
interesting for a number of reasons. We find that there are two
different on-shell $\mathcal{N} = 2$ superalgebras that can
potentially arise as symmetries of these theories. The first of these
is the standard $\mathcal{N} = 2$ superalgebra with the schematic
structure $\{Q^I, Q^J\} \sim \delta ^{IJ} P$; $(I,J) = (1,2)$. In the
case of a flavor $SO(2)$ $R$ symmetry, one can also have a
``mass-deformed'' algebra where the supercharges close on the momentum
as well as the $R$ symmetry generator. Or, schematically, $\{Q^I,
Q^J\} \sim \delta ^{IJ} P + m \epsilon^{IJ} R$. Such mass-deformed
algebras -- though rare in the list of all possible superalgebras --
have been shown to arise as symmetries of three dimensional gauge
theories in a number of previous investigations \cite{SCS, LM,
  AADY2}. In the present work we find that the mass-deformed
$\mathcal{N} = 2$ algebra is the underlying symmetry algebra for the
CSM theory. We find a convenient single particle representation of
this algebra and find that all of the 4-point tree-level amplitudes
can be encoded in superamplitudes (note
that the odd-point amplitudes in the Chern-Simons theory have external
legs which are gauge fields and therefore vanish on-shell). While component amplitudes of $\mathcal{N} \geq 4$ massive CSM theories
have been studied in great detail (for example in\cite{SCS}) the $\mathcal{N} = 2$ CSM theory studied in this paper is distinguished from the class of models investigated in\cite{SCS} by virtue of allowing the matter fields to be in the adjoint representation (which is typically not possible for higher supersymmetry). The component amplitudes obtained in this paper are not known to be obtained by a trivial truncation of the $\mathcal{N} \geq 4$ amplitudes either. Furthermore, to the best of our knowledge the superamplitude presented in this paper for the massive $\mathcal{N} = 2$ theory is the first concrete formulation of a superamplitude for a massive gauge theory in $d = 3$.

In the case of the YMCS theory, we find that the underlying
supersymmetry algebra is the {\it undeformed} one where the supercharges
close on momenta alone. This is to be expected as there is no flavor
symmetry in the bosonic sector of the theory, since the Lagrangian has
only one scalar field. We derive an on-shell representation of this
algebra and use it to obtain constraints on 4-pt. amplitudes (the
on-shell algebra does not constrain the 3-pt. amplitudes). We find
that that the relations among the 4-pt. amplitudes of the YMCS theory
are considerably more complicated than those in the CSM theory. The
root of the complication has to do with the absence of the extra
$SO(2)$ symmetry in the bosonic sector. Nevertheless we are able to compute a
number of these amplitudes and verify that the computed amplitudes are
consistent with the on-shell algebra. Although we do not compute amplitudes with external gauge fields using Feynman diagrams, we are nevertheless
able to deduce the 4-point amplitudes with external gauge fields using the on-shell algebra.

We also propose a BCFW recursion relation for mass-deformed
three-dimensional theories which reduces to the proposal in
\cite{Gang:2010gy} when the mass goes to zero. This recursion relation
involves deforming two external legs of on-shell amplitudes by a complex
parameter $z$. In order for the the recursion relation to be
applicable, the amplitudes must vanish as $z\rightarrow \infty$. We
show that four-point superamplitudes of the CSM theory
have good large-$z$ behavior, so our proposed recursion relation may be
applicable to this theory. However, the proposed relations do not seem to apply to the YMCS theory and  we comment on the relevant issues in the corresponding section of the paper.

Three-dimensional $\mathcal{N}=2$ gauge theories are also interesting
from various other points of view. In particular, they exhibit Seiberg
duality \cite{Aharony:1997gp,Giveon:2008zn,Benini:2011mf},
F-maximization \cite{Jafferis:2010un}, and an F-theorem
\cite{Jafferis:2011zi}. The gravity duals of these theories are also
known and have been studied in
\cite{Martelli:2011qj,Cheon:2011vi,Jafferis:2011zi}. Finally,
three-dimensional $\mathcal{N}=2$ superconformal gauge theories arise
from compactifications of the 6d $(2,0)$ CFT compactified on
3-manifolds \cite{Dimofte:2011ju,Dimofte:2011py,Dimofte:2013iv}. It
would very interesting to make contact with these results from the
point of view of scattering amplitudes.

The structure of this paper is as follows. In section
\ref{lagrangians} we describe some general aspects of the CSM and YMCS
theories whose amplitudes we compute in this paper. We pay special
attention to the derivation of the on-shell supersymmetry algebras in
this section. In particular, we derive the on-shell representation of
the algebra for the YMCS system in some detail following canonical
quantization. This derivation relies on a careful analysis of the implementation  of the Gauss-law constraints on the physical Hilbert space, which is described in some detail.
In section \ref{csamp}, we compute the four-point
amplitudes of the CSM theory and show that they can be encoded in superamplitudes. We also describe the symmetries of the
four-point superamplitudes which we expect to hold for higher-point
superamplitudes. In section \ref{ymcsamp}, we compute various three
and four-point amplitudes of the YMCS theory at tree level and use the
on-shell superalgebra to deduce the remaining 4-pt. amplitudes. We
also comment on the complications that arise when trying to compute
amplitudes with external gauge fields in the YMCS theory using
perturbative techniques. In section \ref{bcfwsection}, we propose a
BCFW recursion relation for mass-deformed 3d theories and discuss its
applicability to the theories studied in this paper. In section
\ref{conclusion}, we present our conclusions and describe some future
directions. In appendix \ref{conventions}, we describe our
conventions, Feynman rules, and various other useful formulae. In
appendix \ref{4fermi} we provide more details about the calculation of
various 4-pt. amplitudes.

\section{Mass-deformed $\mathcal{N} = 2$ gauge theories} \label{lagrangians}

%************************************************************************%
In this section we review some general  aspects of the mass-deformed three-dimensional supersymmetric theories whose scattering amplitudes we  study in this paper. The gauge field which appears in these theories has a Chern-Simons term 
\be S_{CS} = \kappa \int \epsilon^{\mu \nu \rho} \tr(A_\mu
\partial_\nu A_\rho + \frac{2i}{3} A_\mu A_\nu A_\rho) .  \ee 
As is well-known, a Chern-Simons gauge field is not parity invariant
and does not have any propagating degrees of freedom. On the other
hand, a Yang-Mills gauge field respects parity and has one massless
degree of freedom in three dimensions. When taken in conjunction with
the Yang-Mills action, the Chern-Simons term breaks parity and gives
rise to a mass for the three dimensional gluon
\cite{Deser:1981wh,Deser}. There are alternate Lorentz invariant
mass-terms for gluons that one can consider in three dimensions (see
\cite{vpn-thermal, AA-AF} for examples) but they typically lead to
non-local terms in the action. A quadratic mass term for the gauge
field could also arise via the Higgs mechanism, but this would break
gauge invariance. In the present paper we consider the only known
mass-term for a gauge field which is Lorentz invariant, gauge
invariant, and local in three dimensions, namely $S_{CS}$. Note that
$S_{CS}$ admits two different supersymmetric completions leading to
supersymmetric Chern-Simons and Yang-Mills-Chern-Simons theories. In
the first case, the gauge field does not have propagating degrees of
freedom and the physical on-shell degrees of freedom consist of matter
hypermultiplets. In the latter case there are Yang-Mills kinetic terms
and the Chern-Simons term provides a topological mass for the gauge
field (which contributes to the on-shell degrees of freedom). We will
study scattering processes in both the theories while restricting
ourselves to the case of $\mathcal{N} = 2$ supersymmetry.

Before discussing mass-deformed $\mathcal{N}=2$ gauge theories in
greater detail, we briefly review the 3d spinor formalism. The
three-dimensional spinor formalism can be obtained by dimensional
reduction of the four-dimensional spinor
formalism\cite{Lipstein:2012kd}. We begin by writing a 4d null
momentum in bispinor form
\begin{equation}
p^{\alpha\dot{\beta}}=\lambda^{\alpha}\bar{\lambda}^{\dot{\beta}},
\label{4d}
\end{equation}
where $\alpha=1,2$ and $\dot{\beta}=1,2$ are $SU(2)$ indices which
arise from the fact that the Lorentz group is $SO(4)\sim SU(2)_{L}\times SU(2)_{R}$.
When reducing to three dimensions, the distinction between dotted
an undotted indices disappears because the Lorentz group is $SU(2)=\left[SU(2)_{L}\times SU(2)_{R}\right]_{{\normalcolor diagonal}}$. Alternatively, we can reduce to three dimensions by modding out by
translations along a vector field $T^{\alpha\dot{\beta}}$, as described
in \cite{Lipstein:2012kd}. Using the vector field to change dotted indices
to undotted indices in (\ref{4d}) and symmetrizing the indices then gives
\begin{equation}
p^{\alpha\beta}=\lambda^{(\alpha}\bar{\lambda}^{\beta)}.\label{eq:three-dimensionalp}\end{equation}
We symmetrize the indices in order to remove the component of the momentum along
the direction $T^{\alpha\dot{\beta}}$. The resulting momentum
is a $2\times2$ symmetric object, which has three components. 

We denote inner products of the spinors using bracket notation
\[
\left\langle \lambda_{i}\lambda_{j}\right\rangle =\epsilon_{\beta\alpha}\lambda_{i}^{\alpha}\lambda_{j}^{\beta}.
\]
If we square (\ref{eq:three-dimensionalp}), we find
that \begin{equation} \left\langle \lambda\bar{\lambda}\right\rangle
  ^{2}=-4m^{2}{\normalcolor .}\label{eq:mass}\end{equation} Hence, if
the particle is massless, then $\lambda\propto\bar{\lambda}$ and the
momentum can be written in bispinor form as
$p^{\alpha\beta}=\lambda^{\alpha}\lambda^{\beta}$.  More generally,
for a massive particle in three-dimensions, the momentum is given by
(\ref{eq:three-dimensionalp}). Equations (\ref{eq:three-dimensionalp})
and (\ref{eq:mass}) can be summarized as follows
\[
\lambda^{\alpha}\bar{\lambda}^{\beta}=p^{\alpha\beta}-im\epsilon^{\alpha\beta}.
\]   
In particular, $\left\langle \lambda\bar{\lambda}\right\rangle =-2im$.

For later convenience, we will denote $\lambda=u$ and $\bar{\lambda}=-v$. The two spinors $u $ and $v$ are solutions of the free massive Dirac equation and are given in (\ref{spinors})\footnote{An exhaustive list of the properties of these spinors can be found in \cite{SCS}.}. They satisfy \be v^\a u^\b = -p^{\a\b} - im\e^{\a\b},  \ee
where $  P^{\a\b} = - (p_\mu \gamma^\mu C^{-1})^{\b\a}$ is given explicitly by
 \be P^{\a \b} =
P^{\b\a} = \matr{cc}{-p_0 - p_1&p_2\\p_2&-p_0+p_1}.  \ee

%%%%%%%%%%%%%%%%%%%%%%%%%%%%%%%%%%%%%%%%%%%%%%%%%%%%%%%%%%%%%%%%%%%%%%%%%%%
\subsection{$\mathcal{N} = 2$ massive Chern-Simons-Matter (CSM) theory}

The CSM theory is described by the action
\bsp
S_{CSM} &= \kappa \int \epsilon^{\mu \nu \rho} \tr(A_\mu \partial_\nu A_\rho + \frac{2i}{3} A_\mu A_\nu A_\rho) -2\int \tr |D_\mu \Phi |^2 + 2i\int \tr\bar\Psi (D_\mu \gamma^\mu \Psi + m \Psi)\\
&-\frac{2}{\kappa^2}\int \tr\left(|[\Phi,[\Phi^\dagger,\Phi]]+e^2\Phi|^2\right)  + \frac{2i}{\kappa}\int \tr([\Phi^\dagger, \Phi][\bar \Psi, \Psi] + 2[\bar \Psi, \Phi][\Phi^\dagger, \Psi]),
\end{split}\label{n=2csm}
\ee
where,
 \be
 \kappa  = \frac{k}{4\pi}, \hspace{.3cm} m = e^2/\kappa.
 \ee
Note that the Chern-Simons term is odd under parity, so the theory is
not parity invariant. The parameter $k$ is the Chern-Simons level. The
matter couples to the gauge field with coupling constant
$1/\sqrt{k}$. The parameter $e$ sets the mass-scale in the
superpotential. Even though it is a dimension-full number, it does not
run in the super-renormalizable theory and can be regarded as a
parameter of the theory. Taking the mass to zero or infinity while
holding the coupling $1/\sqrt{k}$ constant corresponds to taking $e$
to zero or infinity. In the massless limit this theory reduces to a
conformal $\mathcal{N}=2$ Chern-Simons-matter theory. In the
infinitely massive limit, the theory reduces to a pure Chern-Simons
theory with no propagating degrees of freedom. The conventions
underlying the above action assume that all the fields are in the
adjoint representation of the gauge group. Furthermore, we assume the
generators of the gauge group $t^a$ (which we can take to be $SU(N)$)
to be Hermitian. We then have
\bsp\label{sunstuff} &A =
A^at^a,\quad \Phi = \Phi^at^a,\quad \Psi = \Psi^a t^a,\\ 
&\tr(t^at^b) = \frac{1}{2}
\delta ^{ab},\quad [t^a, t^b] = if^{abc}t^c,\quad D_\mu = \partial_\mu -
i[A_\mu,\hspace{.1cm}].
\end{split}
\ee
In terms of real variables,
\be
\Phi  = \frac{1}{\sqrt{2}}(\Phi ^1 + i \Phi^2), \quad
\Psi = \frac{1}{\sqrt{2}}(\Psi ^1 + i \Psi^2),
\ee
where $\Phi^i$ and $\Psi^i$ are real and Majorana respectively.

We can immediately see that the free (Abelian) part of the action is invariant under
\bsp
&\delta _{\bar \epsilon} \Phi = \bar \epsilon \Psi, \hspace{.2cm} \delta_\epsilon \Phi^\dagger =  \bar\Psi\epsilon,\\
&\delta _\epsilon \Psi = +i(\partial^\mu \gamma_\mu - m)\Phi\epsilon, \hspace{.2cm} \delta_{\bar \epsilon}\bar \Psi = -i\bar \epsilon(\partial^\mu \gamma_\mu + m)\Phi^\dagger,
\end{split} \label{csmfree}
\ee
where $\delta _\epsilon = [\bar Q\epsilon, \hspace{.1cm}]$, $\delta _{\bar\epsilon} = [\bar \epsilon Q,\hspace{.1cm}]$. All other supersymmetry variations vanish. In the non-Abelian / interacting theory, the SUSY variation of the scalar fields remains as above, but the variations of the fermions and the gauge fields are given by
\bsp
\delta _\epsilon \Psi &= \left(i(\partial^\mu \gamma_\mu - m)\Phi - \frac{i}{\kappa}[\Phi,[\Phi^\dagger, \Phi]]\right)\epsilon,\\
\delta_\epsilon A_\mu &= -\frac{i}{\kappa} [\Phi, \bar \Psi\gamma_\mu\epsilon].
\end{split}
\ee
The $\delta _{\bar \epsilon}$ variations in the non-Abelian case can be obtained from the ones given above by conjugation. The fundamental anti-commutation relation between the supercharges is
\be\label{def-alg}
\{Q^{\beta J}, Q^{\alpha I}\} = \frac{1}{2}\left( P^{\alpha \beta }\delta ^{IJ} + m \epsilon^{\beta \alpha }\epsilon^{JI}R\right),
\ee
where $R$ is the $SO(2)=U(1)$ symmetry generator which rotates $(\Phi^1,\Phi^2)$ and $(\Psi^1,\Psi^2)$.

For the mass-deformed Chern-Simons theory, the on-shell asymptotic states are those of the complex scalar $\Phi$ and fermion $\Psi$. In our  notation, the asymptotic momentum-space states of $\Phi  $ and $\Psi$ are denoted $|a_+\rangle$ and $|\chi _+\rangle$ respectively. Using the mode expansions for these fields, which are given by (\ref{ymcs-mode}), in the supersymmetry algebra (\ref{csmfree}), we see that the supersymmetry variations of the on-shell states are given by 

\bsp &
Q_I|\Phi_1\rangle = \frac{1}{2}v|\chi_I\rangle,\\ & Q_I|\Phi_2\rangle =
\frac{1}{2}v\epsilon^{IJ}|\chi_J\rangle,\\ & Q_I|\chi_J\rangle =
\frac{1}{2}\delta_{IJ}u|\Phi_1\rangle +
\frac{1}{2}\epsilon^{IJ}u|\Phi_2\rangle,
\end{split}\label{def-n-2}
\ee
where $u$ and $v$ are spinors defined in (\ref{spinors})\footnote{For a detailed discussion of the on-shell representation of three dimensional massive $\mathcal{N} \geq 4$ superalgebras, we refer to \cite{SCS}.}. 

We can express these transformations in a way that makes the $U(1)$ R-symmetry of the theory manifest by forming complex combinations of the fields
and supercharges
\be\label{pmconvention}
a_\pm =
\frac{1}{\sqrt{2}}(\Phi_1 \pm i\Phi_2),\quad \chi_\pm =
\frac{1}{\sqrt{2}}(\Psi_1 \pm i\Psi_2),\quad Q_\pm= \frac{1}{\sqrt{2}}(Q_1 \pm i Q_2).
\ee 
We then obtain
\bsp &Q_+|a_+\rangle =
\frac{1}{\sqrt{2}}v|\chi_+\rangle, \hspace{.3cm} Q_+|\chi_-\rangle =
\frac{1}{\sqrt{2}}u|a_-\rangle,\\ &Q_-|a_-\rangle =
\frac{1}{\sqrt{2}}v|\chi_-\rangle, \hspace{.3cm} Q_-|\chi_+\rangle =
\frac{1}{\sqrt{2}}u|a_+\rangle,\\ &Q_-|a_+\rangle = Q_+|\chi_+\rangle
= Q_+|a_-\rangle = Q_-|\chi_-\rangle = 0.
\end{split}\label{n=2c}
\ee

It is important to emphasize that the superalgebra (\ref{def-alg}) is
a non-central extension of the standard $\mathcal{N} = 2$
superalgebra. In particular, the anticommutator of the charges does
not close onto the momentum generator alone, as it also involves the
$R$ symmetry generator as part of the fundamental supersymmetry
algebra.  Such mass-deformed algebras frequently arise in the context
of three dimensional gauge theories with mass-gaps; in particular in
$\mathcal{N} \geq 4$ Chern-Simons-Matter theories \cite{SCS}. It is
instructive to see how the algebra described above is embedded in the
supersymmetry algebra of the massive $\mathcal{N} = 6$ theory. In the
notation of \cite{SCS}, the matter content of $\mathcal{N} = 5,6$ CSM
theories is given by scalars $\phi_a, \tilde \phi_{\dot a}$ and
fermions $\psi_{\dot a}, \tilde \psi_a$, where $a,\dot a$ are two
different $SU(2)$ indices. The fields denoted by tilde are part of the
twisted hypermultiplets, while those without the tildes form the
untwisted hypermultiplets. In the case of $\mathcal{N}=6$
supersymmetry, one has - as part of the full supersymmetry algebra -
supercharges $Q^\pm_\alpha $ that transform scalars and fermions
belonging to the twisted and untwisted hypermultiplets to each other,
while acting trivially on the $SU(2)$ indices (see the discussion in
section-2.4 of \cite{SCS}). These supercharges (for a fixed value of
the $SU(2)$ index) generate the massive $\mathcal{N} = 2$ algebra
considered here\footnote{There are presumably other distinct
  embeddings of the $\mathcal{N} = 2$ superalgebra in the larger
  $\mathcal{N} = 6$ superalgebra as well.}. It should be noted,
however, that the theory considered here is distinguished from the
class of models studied in \cite{SCS} by virtue of all the matter
fields being in the adjoint representation, which, typically is not
possible for $\mathcal{N} = 4$ and higher supersymmetry.

We also note that just as
the supersymmetric Chern-Simons theories are not known to be obtained as the dimensional
reduction of higher dimensional gauge theories, this massive
superalgebra is {\it not} what one obtains by the dimensional reduction of
the free $\mathcal{N} = 1$ theory in four dimensions.  In fact, if one
takes the massive $\mathcal{N}=1$ $d=4$ free action given by
\be
S_{d=4} = -\frac{1}{2}\int_{R^4} \left(\partial_\mu \Phi_I
\partial^\mu \Phi_I + m^2\Phi_I\Phi_I +i\bar{\Psi }\G_\mu \partial
^\mu \Psi+ im\bar{\Psi}\Psi\right), \ee which is invariant under \bsp
& \delta_\a \Psi = \frac{1}{2}(\G^\m\partial_\m - m)\Phi_1 \alpha +
\frac{i}{2} (\G^5\G^\m \partial_\mu + m \G^5)\Phi_2\a,\\ &\delta_\a
\Phi_1 = \frac{i}{2}\bar \a \Psi, \hspace{.3cm} \delta_\a \Phi_2 =
\frac{1}{2} \bar \a \G^5 \Psi,
\end{split}
\ee 
it is easy to see that 
the algebra closes on only the momentum generators without
any extensions \be [\delta_ \b, \delta _\a]\Phi_I = \frac{i}{2}(\bar
\a \G^\m \b)\partial_\mu \Phi_I.  \ee 

The algebra retains this
standard form even after dimensional reduction to $d=3$, however the
fermion mass-term in $d=3$ derived from the $SO(1,3)$-invariant four
dimensional mass-term would be  given in the three dimensional
notation by
 $\int (\bar
\Psi_1\Psi_1 - \bar\Psi_2\Psi_2)$. 
This is different from the term we
have in (\ref{n=2csm}) where the mass terms for both the fermions have the
same sign.

In other words, in $d=3$ we can choose between two different fermion
mass-terms \be M_1 = \int (\bar \Psi_1\Psi_1 -
\bar\Psi_2\Psi_2), \hspace{.2cm}\text{or} \hspace{.2cm} M_2 = \int
\bar{\Psi}_I\Psi_I.  \ee The choice $M_1 $ -- the parity conserving
option -- leads to the the standard $\mathcal{N}=2$ algebra without
extensions while $M_2$ leads to a mass-deformed algebra and violates
parity. However the Chern-Simons term, which is present in the gauge
theories we study, violates parity. Thus it is natural that the
fermionic mass terms resulting from the supersymmetric completion of
the Chern-Simons term violate parity as well. It is apparently this
interplay between the parity invariance of the theory and the
fermionic mass term that leads to the massive nature of the on-shell
algebra in this case.

%%%%%%%%%%%%%%%%%%%%%%%%%%%%%%%%%%%%%%%%%%%%%%%%%%%%%%%%%%%%%%%%%%%%%%
\subsection{$\mathcal{N} = 2$ Yang-Mills-Chern-Simons (YMCS) theory}\label{ymcsformal}

The second theory of relevance to this paper is the well-known
$\mathcal{N} = 2$ YMCS theory described off-shell by the action
$S_{YMCS} = S_{YM} + S_{CS}$ where\footnote{We have taken the trace
  using (\ref{sunstuff}) and also chosen to rescale the fields by the coupling
constant, in comparison to (\ref{n=2csm}).}
\bsp 
S_{YM} &= \frac{1}{e^2} \int\Biggl[ -\frac{1}{4} F^a_{\mu \nu}F^{a\mu \nu} -
\frac{1}{2}  D_\mu \Phi ^a D^\mu \Phi^a + \frac{1}{2}
F^aF^a \\ 
& +\frac{i}{2} \bar \Psi ^a_I\gamma ^\mu D_\mu\Psi ^a_I 
+\frac{i}{2}f^{abc}  \epsilon_{AB}\bar \Psi _A^a
 \Phi^b\Psi _B^c \Biggr],\\ 
S_{CS} &= \frac{m}{2e^2}\int\Biggl[
\epsilon^{\mu \nu \rho} A^a_\mu \partial_\nu A^a_\rho 
-\frac{1}{3}f^{abc}\epsilon^{\mu \nu \rho}  A^a_\mu A^b_\nu
A^c_\rho + i\bar\Psi^a _I\Psi_I^a +
2F^a\Phi^a\Biggr].\label{ymcs}
\end{split}
\ee
Here $m = \frac{ke^2}{4\pi}$ where $k$ is the Chern-Simons level,
$A=1,2$ is an $SO(2)$ index, the scalar field $\Phi$ and auxiliary
field $F$ are real, and the fermions are Majorana, so that $\bar\Psi_A
\equiv \Psi_A^\dag \g^0 =\Psi_A^T C$, where the charge conjugation
matrix $C=\g^0$. For further details about our conventions, see
appendix \ref{conventions}. Note that the matter fields $\Phi$ and
$\Psi$ have mass dimension $1$ and $3/2$. If they are rescaled by a
factor of $e$, i.e. if $(\Phi,\Psi)\rightarrow e (\Phi,\Psi)$, then
they will have mass dimensions $1/2$ and $1$. $F$ is an auxiliary
field whose elimination generates the standard quadratic mass term of
the real scalar $\Phi$, while the Chern-Simons term gives a
topological mass to the gauge field. Both the Chern-Simons term and
the fermionic mass term are odd under parity, so the theory is not
parity invariant when the mass is nonzero. Taking the mass to zero or
infinity while holding the Yang-Mills coupling constant corresponds to
taking $k$ to zero or infinity. In the massless limit, the theory
reduces to $\mathcal{N}=2$ Yang-Mills theory and parity is
restored. In the infinitely massive limit, the theory reduces to a
pure Chern-Simons theory. What is not transparent from the action
given above, but is nevertheless true, is that the asymptotic physical
states also involve a second massive scalar with the same mass. This
scalar is nothing but the physical gauge invariant degree of freedom
encoded in the gauge field.

Note that for the free theory, the supersymmetry
transformation laws are:
\bsp\label{linsusy}
\delta_\eta A_\mu = -\frac{i}{2}(\bar{\eta}_I \gamma_\mu \Psi_I), \hspace{.3cm} \delta_\eta \Phi = \frac{1}{2}\bar{\eta}_I \Psi_J \e_{IJ},\\
\delta_\eta\Psi_I = \frac{1}{4}\gamma ^{\mu \nu }F_{\mu \nu} \eta_I - \frac{i}{2}(\gamma ^\mu \partial_\mu -m)\e_{IJ}\eta_J \Phi.
\end{split}
\ee
The anti-commutator of the supercharges in these off-shell transformation laws closes onto the momentum operator alone
\be
\{Q^\a_I, Q^\b_J\} = -\frac{1}{2}\delta_{IJ}P^{\a\b}.
%\label{undeformed}
\ee
Note that there is no $U(1)$ extension as there was for the
superalgebra in the mass-deformed Chern-Simons theory. This is a
consequence of the fact that the YMCS theory, while enjoying a $SO(2)$
R-symmetry which rotates the two fermionic fields in the theory, does
not have a corresponding symmetry acting on the two bosonic fields,
i.e. the scalar and gauge field. Indeed, we will find that the
on-shell amplitudes of the YMCS theory exhibit $SO(2)$ R-symmetry in
the fermionic sector. We do note, however, that both the algebras
collapse to the same massless algebra when $k$ in the YMCS theory and $e$
in the CSM model are set to zero, which is consistent with the
four-point amplitudes of undeformed three-dimensional SYM exhibiting
an enhanced $SO(2)$ symmetry\cite{AADY1, Lipstein:2012kd}.

We now focus on the on-shell superalgebra for this theory.
Assuming that the on-shell degree of freedom associated with the YMCS
gauge field corresponds to a massive scalar field (which we will
justify shortly) and that the superalgebra in (\ref{linsusy}) is
realized on the single particle asymptotic states, the
transformation laws for the scattering states can be taken to be
\bsp\label{undeformed}
&Q_I |a_1\rangle = \frac{1}{2}u|\chi _I \rangle, \hspace{.2cm} Q_I |a_2\rangle  = \frac{1}{2}\e_{IJ} v|\chi_J\rangle,\\
&Q_J|\chi_I\rangle = \frac{1}{2}\delta_{IJ}v|a_1\rangle - \frac{1}{2}\e_{IJ}u|a_2\rangle.
\end{split}
\ee
We denote the massive scalar corresponding to the gauge field by $a_1$.

One can give an argument in favor of the algebra above being the
appropriate one for the YMCS theory as follows. If one starts with the
part of the algebra involving the variation of $a_2$, namely
$Q_I|a_2\rangle = \frac{1}{2}\e_{IJ} \tau |\chi_J\rangle$, there is an
ambiguity about what the spinor $\tau $ is. This ambiguity can be
resolved by applying the oscillator expansion of the fields to the
off-shell transformation $ \delta \Phi = \frac{1}{2}\bar{\eta}_I
\Psi_J \e_{IJ}$. In our convention, this fixes $\tau = v$. Once this
is fixed, the closure of the algebra on $a_2$ fixes the transformation
properties $Q_1|\chi_2\rangle = + u/2|a_2\rangle $ and
$Q_2|\chi_1\rangle = - u/2|a_2\rangle$. With this part of the on-shell
supersymmetry algebra determined, one can make the following ansatz
for the supersymmetry algebra
\bsp
&Q_I |a_1\rangle = \frac{1}{2}\omega|\chi _I \rangle, \hspace{.2cm} Q_I |a_2\rangle  = \frac{1}{2}\e_{IJ} v|\chi_J\rangle,\\
&Q_J|\chi_I\rangle = \frac{1}{2}\delta_{IJ}\tilde{\omega}|a_1\rangle - \frac{1}{2}\e_{IJ}u|a_2\rangle,
\end{split}
\ee
assuming that the realization is linear in the fields and that the
$SO(2)$ covariance of the fermionic degrees of freedom is
respected. The unknown quantities are the spinors $\omega $ and
$\tilde \omega $.  Closure of the algebra on $a_1$ requires $\omega
^{\{\a}\tilde\omega^{\b\}} = -2P^{\a\b}$. The solution to this
equation is given by $\{\omega,\tilde\omega\} = \{u,v\}$ or
$\{\omega,\tilde\omega\} = \{v,u\}$.  Furthermore, requiring that
there be no $U(1)$ extension to the algebra requires $\omega = u$ and
$\tilde \omega = v$.  Thus, given a convention of the oscillator
expansion of the fermion fields, the on-shell algebra is unambiguously
determined.

Comparing this with (\ref{def-n-2}), we see that main difference
between the two sets of transformations is that the spinors appearing
on the RHS of the transformation laws of the scalars above are
conjugates of each other. The two spinors were same in the
transformation laws for the CSM theory. The differences in the two
realizations have to do with whether or not the algebra is
mass-deformed.

Rather than rely on the argument above alone, it is instructive to
derive (\ref{undeformed}) using the methods of canonical quantization.
To this end we revert to a Hamiltonian framework and set $A_0 = 0 $. We
define the complex combination $A = \frac{1}{2}(A_1 + i A_2)$ (and its
conjugate relation) for the gauge potentials. Due to the
non-commutativity induced by the Chern-Simons term on the components
of the electric field the Gauss law constraints can be shown to be
satisfied by wave functions of the form\cite{AAVPN} 
\be \Omega =
\exp\left[\frac{k}{2}\left( S_{WZW}(M^\dagger) - S_{WZW}(M)\right)\right] \Lambda,
\ee 
where $M$ is a complex matrix related to the gauge potential as $A
= - \partial M M^{-1}$ that transforms under time independent local
gauge transformations as $M \rightarrow UM$, where $U$ is an element
of the gauge group.  $S_{WZW}$ is a Wess-Zumino-Witten functional
defined over the spatial manifold \cite{AAVPN}. The Gauss law
constraint can be translated into the following condition on
$\Lambda$ 
\be (D\frac{\delta}{\delta A} + \bar D \frac{\delta}{\delta
  \bar A})^a + f^{amn}(-i\bar \Psi ^m_I\Psi ^n_I + \Phi^m \frac{\delta
}{\delta \Phi^n})\Lambda = 0.  \ee 
Clearly any wave functional
$\Lambda$ that is a gauge invariant combination of the gauge and
matter fields satisfies this constraint.

Now, to derive the on-shell supersymmetry transformation law, our strategy would be to express the quadratic part of the supercharges in terms of the canonical variables followed by a dualization of the gauge field into a scalar. We can then read off the on-shell supersymmetry transformation by looking at the action of the dualized supercharge on the dynamical fields in momentum space. To avoid the ambiguity associated with the fermionic fields and their canonical momenta in a real representation for the three dimensional Dirac matrices, we take the gamma matrices to be $\gamma ^\m = \{i\sigma ^3, \sigma ^1, \sigma ^2\} $ for the purposes of this discussion (everywhere else in the paper we shall continue to use the real $\g$ matrices mentioned previously). The fermions can be taken to be 
$\Psi = \matr{c}{\psi \\ \psi^*}$ with $\psi $ and $\psi ^*$ being
canonically conjugate. The quadratic part of the top component of the
$\mathcal{N} = 2$ supercharge in this notation (the bottom component
can simply be obtained by Hermitian conjugation) can be written as
\bsp
q_I &= ie\int   \psi^{ a \dagger}_I \frac{\delta }{\delta A^a} - \frac{1}{e} \int \psi ^a_I B^a + e\epsilon_{IJ}\int \psi ^a_J (\Pi^a_{\Phi} + \frac{im}{e^2}\Phi^a)- \frac{2i}{e} \epsilon_{IJ}\int \psi^{a\dagger}_J (\bar D \Phi)^a.
\end{split}
\ee
This charge, derived from the action, acts on the wave function $\Omega$. $\Omega $ and $\Lambda$ differ by a pure phase, so their norms are the same. However the physical observables acting  on $\Lambda $ differ from those acting on $\Omega $ by a unitary transformation. The charge acting on $\Lambda $ = $q' _I= \Omega^\dagger q \Omega $ \cite{AAVPN}. The effect of this unitary transformation is to replace 
\be
\frac{\delta }{\delta A^a} \rightarrow \frac{\delta }{\delta A^a} + \frac{m}{e^2}(\bar A^a - \bar a^a), \hspace{.3cm} \bar a   = \bar \p M M^{-1}.
\ee
This extra term generated by the unitary transformation is what generates an effective mass-term for the gauge field in the Hamiltonian obtained from the supercharge above\cite{AAVPN}.

We can now dualize the gauge field by expressing $M = e^\theta$ and
retaining terms to linear order in $\theta$. This gives (after
dropping the color indices, as we are only interested in the abelian
theory) $A = -\p\theta$, $\bar A = +\bar \p\bar\theta$, $a = \p\bar
\theta$ and $\bar a = -\bar \p\theta$. The real part of $\theta $
is related to the physical gauge-invariant on-shell degree of freedom
$\Phi_H$ as \cite{Deser:1981wh, AAVPN, KKNCS}
\be \theta + \bar \theta
= \frac{1}{\sqrt{-\p\bar\p}}\Phi_H.  \ee On gauge invariant wave
functionals, the dualized supercharge can now be written as
\be\label{dualq} q'_I = ie \omega_I(\frac{\delta}{\delta \Phi_H} +
\frac{im}{e^2}\Phi_H) + \frac{2i}{e}\omega^\dagger_I\bar\p\Phi_H +
e\epsilon_{IJ}\int \psi _J (\Pi_{\Phi} + \frac{im}{e^2}\Phi)-
\frac{2i}{e} \epsilon_{IJ}\int \psi^{\dagger}_J (\bar \p \Phi), \ee
where \be \omega _I = -ie^{i\a}\psi^\dagger, \hspace{.3cm} e^{i\a} =
\sqrt{\bar \p/\p} \equiv \sqrt{\bar p/p}.  \ee 
The momenta $p$ and $\bar p$ appearing in the last term above are the complex
combinations of the spatial components of the three-momentum. It is
important to note that the fermionic variable multiplying the momentum
for the dual scalar has to be identified as the top component of a
fermionic field (in our case $\omega$) so that the SUSY variation of
$\Phi_H$ can be written in a Lorentz invariant form in the two
component notation i.e. $\delta \Phi_H \sim \bar \epsilon \rho$ for
some Majorana fermion $\rho$. In our case the dualization dictates
that the top component of $\rho$ is $\omega $. Crucially for our
purposes, it can be readily seen from the Hamiltonian obtained
from $q'_I$ that the Hamiltonian for $\omega $ has the opposite sign
for the mass term as that of $\psi$. Or in other words, the spinors
appearing with the positive (negative) frequency parts of the mode
expansion of $\psi$ can be identified with those
associated with the negative (positive) parts of $\omega $. Since the
SUSY variations involving the on-shell fields $a_1 \equiv \Phi_H$ and
$a_2 \equiv \Phi$ involve fermions with the opposite mass terms, the
spinors appearing in the momentum space realization of these
transformations are conjugate to each other.  Reverting back to our
real conventions for the $\g$ matrices, we see that (\ref{undeformed})
can now be justified based on the grounds of canonical quantization.

\section{Mass-deformed Chern-Simons amplitudes} \label{csamp}

In this section, we will describe the scattering amplitudes of the CSM
theory (\ref{n=2csm}). Since the Chern-Simons gauge field has no propagating degrees
of freedom, scattering amplitudes with at least one external gauge
field vanish. This implies that all odd-point amplitudes vanish, so
the first nontrivial amplitudes occur at four-point. In the next two
subsections, we will compute the 4-point amplitudes and show that they
can be encoded in superamplitudes. We also describe the symmetries of
these superamplitudes.

\subsection{4-point amplitudes} \label{amp4}

All of the the four point amplitudes of the CSM theory are related to
each other by the supersymmetry algebra in (\ref{n=2c}). Hence, there
are only two independent amplitudes involving four legs. With the definitions of the complex combinations of the real degrees of freedom
described in (\ref{pmconvention}), we get the following relations
between the color ordered four particle amplitudes

\bsp\label{bbff} &\langle \chi_+ \chi_- a_+a_-\rangle =
+\frac{\tprods{\bar 4}{ 1}}{\tprods{\bar 2}{\bar 4}}\langle a_+a_-a_+a_-\rangle,\quad
\langle \chi_+ \chi_- a_-a_+\rangle =
+\frac{\tprods{\bar 3}{ 1}}{\tprods{\bar 2}{\bar 3}}\langle a_+a_-a_-a_+\rangle,\\
&\langle a_+\chi_- \chi_+ a_-\rangle =
+\frac{\tprods{ 3}{\bar 4}}{\tprods{\bar 2}{\bar 4}}\langle a_+a_-a_+a_-\rangle,\quad
\langle a_+\chi_- a_-\chi_+ \rangle =
+\frac{\tprods{\bar 3}{ 4}}{\tprods{\bar 3}{\bar 2}}\langle a_+a_-a_-a_+\rangle,\\
&\langle \chi_+ a_+\chi_-a_- \rangle =
+\frac{\tprods{\bar 4}{ 1}}{\tprods{\bar 3}{\bar 4}}\langle a_+a_+a_-a_-\rangle,\quad
\langle a_+ \chi_+\chi_-a_- \rangle =
+\frac{\tprods{\bar 4}{ 2}}{\tprods{\bar 3}{\bar 4}}\langle a_+a_+a_-a_-\rangle,\\
&\langle \chi_+ a_-\chi_- a_+\rangle =
+\frac{\tprods{ 1}{ \bar 2}}{\tprods{\bar 2}{\bar 3}}\langle a_+a_-a_-a_+\rangle,\quad
\langle a_+a_-\chi_- \chi_+ \rangle =
+\frac{\tprods{\bar 2}{ 4}}{\tprods{\bar 2}{\bar 3}}\langle a_+a_-a_-a_+\rangle,\\
&\langle a_+a_- \chi_+ \chi_-\rangle =
+\frac{\tprods{ 3}{\bar 2}}{\tprods{\bar 2}{\bar 4}}\langle a_+a_-a_+a_-\rangle,\quad
\langle \chi_+a_-a_+ \chi_- \rangle =
+\frac{\tprods{ 1}{\bar 2}}{\tprods{\bar 2}{\bar 4}}\langle a_+a_-a_+a_-\rangle,\\
&\langle a_+\chi_+a_- \chi_- \rangle =
+\frac{\tprods{ 2}{\bar 3}}{\tprods{\bar 3}{\bar 4}}\langle a_+a_+a_-a_-\rangle,\quad
\langle \chi_+a_+a_- \chi_- \rangle =
+\frac{\tprods{ 1}{\bar 3}}{\tprods{\bar 3}{\bar 4}}\langle a_+a_+a_-a_-\rangle.\\
\end{split}
\ee
The three independent four-fermion amplitudes are related to the other amplitudes as
\bsp\label{ffff} &\langle \chi_+\chi_+\chi_-\chi_-\rangle
= +\frac{\tprods{ 2}{ 1}}{\tprods{ \bar 4}{2}}\langle
a_+\chi_+\chi_-a_-\rangle  =  +\frac{\tprods{ 2}{ 1}}{\tprods{\bar
    3}{\bar 4}}\langle
a_+a_+a_-a_-\rangle,\\
&\langle \chi_+\chi_-\chi_+\chi_-\rangle =
+\frac{\tprods{  1}{ 3}}{\tprods{  3}{\bar 2}}\langle
a_+a_-\chi_+\chi_-\rangle = +\frac{\tprods{  1}{ 3}}{\tprods{\bar
    2}{\bar 4}}\langle
a_+a_-a_+a_-\rangle,  \\
&\langle \chi_+\chi_-\chi_-\chi_+\rangle =
+\frac{\tprods{ 4}{ 1}}{\tprods{\bar 2}{ 4}}\langle
a_+a_-\chi_-\chi_+\rangle =  +\frac{\tprods{ 4}{ 1}}{\tprods{\bar
    2}{\bar 3}}\langle
a_+a_-a_-a_+\rangle.
\end{split}
\ee

In appendix \ref{4fermi} we compute 4-fermion amplitudes and find
\[
\langle \chi_+\chi_+\chi_-\chi_-\rangle_{CSM}= 2i\frac{\la 4 3\ra\la 4  2\ra}{\la 4\bar 1\ra} 
\]
\begin{equation}
\langle \chi_+\chi_-\chi_+\chi_-\rangle_{CSM} = 2i\left\langle 24\right\rangle \left(\frac{\left\langle 41\right\rangle \left\langle 4\bar{1}\right\rangle -\left\langle 43\right\rangle \left\langle 4\bar{3}\right\rangle }{\left\langle 4\bar{1}\right\rangle \left\langle 4\bar{3}\right\rangle }\right).
\label{4fermics}
\ee
Using these amplitudes, all the other 4-pt. amplitudes are determined from the above relations and cyclic permutations. 

\subsection{Superamplitudes} 

The natural question to ask is if these relations among the 4-point amplitudes obtained in the previous subsection can be derived from  superamplitudes. To the best of our knowledge no superamplitude is known for any massive $CSM$ model  so far. However, since the superalgebra and the kinematics constraining the $S$-matrix of the massive three dimensional theories can be thought of as dimensional reductions  of the four dimensional quantities, it is natural to expect that some of the known results for four dimensional $SYM$ theories can be reduced to get massive three dimensional superamplitudes.
In fact, one can define two types of superamplitudes, which are analogous to the ``$\Phi-\Psi$" formalism and the   ``$\Phi-\Phi^\dagger$" formalisms used to obtain superamplitudes for  4d  super-Yang-Mills theories with $\mathcal{N}<4$ supersymmetry \cite{Elvang:2011fx}. In the ``$\Phi-\Psi$" formalism, the superamplitudes can be expressed in terms of supercharges so one can in principle apply super-BCFW recursion relations to these amplitudes. On the other hand, the $SO(2)$ R-symmetry of the theory is not manifest in the ``$\Phi-\Psi$" formalism, so the on-shell superalgebra obtained from the supercharges which act on the superamplitudes does not have a central extension. In the ``$\Phi-\Phi^\dagger$" formalism, the superamplitudes are not expressed in terms of supercharges, but the superamplitudes and superfields transform covariantly under the $U(1)$ R-symmetry. We describe the ``$\Phi-\Psi$" and ``$\Phi-\Phi^\dagger$" formalisms in greater detail below, and we describe super-BCFW for 3d mass-deformed theories in section \ref{bcfwsection}.        

\subsubsection{$\Phi-\Psi$ Formalism} \label{superamp}

We introduce the on-shell superfields
\begin{equation}
\Phi=a_{+}+\bar{\eta}\chi_{+},\,\,\,\Psi=\chi_{-}+\eta a_{-},
\label{superfields}
\end{equation}
where $\eta$ is a complex Grassmann variable. The 4-pt superamplitudes can then be written in terms of a supermomentum delta function
\begin{equation}
\mathcal{A}_{4}=\Omega\,\delta^{3}(P)\delta^{2}(Q),
\label{4ptsuperamp}
\end{equation}
where $\Omega$ is a prefactor and
\begin{equation}
P^{\alpha\beta}=\sum_{i=1}^{4}\lambda_{i}^{(\alpha}\bar{\lambda}_{i}^{\beta)},\,\,\, Q^{\alpha}=\sum_{i=1}^{4}\left(\lambda_{i}^{\alpha}\bar{\eta}_{i}+\bar{\lambda}_{i}^{\alpha}\eta_{i}\right),\,\,\,\delta^{2}(Q)=Q^{\alpha}Q_{\alpha}.
\label{supermomentum}
\end{equation}

The 4-pt. superamplitudes of
the CSM theory are given by

\[
{\cal A}_{\Phi\Phi\Psi\Psi}=\frac{\left\langle 2 4\right\rangle }{\left\langle \bar{3}2\right\rangle }\delta^{3}(P)\delta^{2}(Q),
\]
\begin{equation}
{\cal A}_{\Phi\Psi\Phi\Psi}=\frac{\left\langle 41\right\rangle \left\langle 4\bar{1}\right\rangle -\left\langle 43\right\rangle \left\langle 4\bar{3}\right\rangle }{\left\langle 1\bar{2}\right\rangle \left\langle 4\bar{1}\right\rangle } \delta^{3}(P)\delta^{2}(Q),
\label{eq:cssuperamp}\end{equation}
where we have ignored the numerical prefactor $2 i$.  Note that the superamplitudes encode the scattering of all component fields. In particular, the component amplitudes correspond to the coefficients of the Taylor expansions of the superamplitudes in the fermionic variables. For example, the coefficient of $\bar{\eta}_{1}\bar{\eta}_{2}$ in the Taylor expansion of $A_{\Phi\Phi\Psi\Psi}$ is 
\[
\left\langle 12\right\rangle \frac{\left\langle 24\right\rangle }{\left\langle \bar{3}2\right\rangle }=\left\langle 12\right\rangle \frac{\left\langle 34\right\rangle \left\langle 24\right\rangle }{\left\langle 34\right\rangle \left\langle \bar{3}2\right\rangle }=\frac{\left\langle 34\right\rangle \left\langle 42\right\rangle }{\left\langle \bar{1}4\right\rangle }
\]
where we noted that $\left\langle 34\right\rangle \left\langle \bar{3}2\right\rangle =-\left\langle \bar{1}4\right\rangle \left\langle 12\right\rangle$. This indeed matches our result for the $\left\langle \chi_{+}\chi_{+}\chi_{-}\chi_{-}\right\rangle$  amplitude in (\ref{4fermics}). Using a similar analysis, one sees that the $\left\langle\chi_{+}\chi_{-}\chi_{+}\chi_{-}\right\rangle$  amplitude in (\ref{4fermics}) corresponds to the $\bar{\eta}_{1}\bar{\eta}_{3}$ component of $A_{\Phi\Psi\Phi\Psi}$. Furthermore, it is easy to reproduce the relations in (\ref{bbff}) and (\ref{ffff}). Hence, the superamplitudes $A_{\Phi\Phi\Psi\Psi}$ and $A_{\Phi\Psi\Phi\Psi}$ encode all the 4-point component amplitudes in the mass-deformed Chern-Simons theory.    

In addition to the supercharge defined in (\ref{supermomentum}), we can also define the following supercharge which annihilates the 4-point superamplitudes
\begin{equation}
\bar{Q}^{\alpha}=\sum_{i=1}^{n}\left(\bar{\lambda}_{i}^{\alpha}\frac{\partial}{\partial\bar{\eta}_{i}}+\lambda_{i}^{\alpha}\frac{\partial}{\partial\eta_{i}}\right){\normalcolor .}\label{eq:Q2}\end{equation}
The superalgebra which acts on the superamplitudes of the CSM theory is given by
\[
\left\{ \bar{Q}^{\alpha},Q^{\beta}\right\}  \propto P^{\alpha\beta}.\] 
Note that the superalgebra does not have a central extension, like the one in (\ref{def-alg}). This is because the $SO(2) \sim U(1)$ R-symmetry of the theory is not manifest in the ``$\Phi-\Psi$" formalism.  The charges acting on the superamplitude and superfields can be regarded as the subset of charges in (\ref{def-alg}) that carry the same $SO(2)$ index, and thus do not have the central term in their anti-commutator.
In particular, the $U(1)$ R-symmetry acts on the on the $\left(\lambda,\eta\right)$ variables as follows: 
\[
\left(\lambda,\eta\right)\rightarrow\alpha\left(\lambda,\eta\right),\,\,\,\left(\bar{\lambda},\bar{\eta}\right)\rightarrow\alpha^{-1}\left(\bar{\lambda},\bar{\eta}\right)
\]
where $\alpha\in U(1)$. Under this symmetry, fields of helicity $h$
should be multiplied by $\alpha^{2h}$, however the superfields in
(\ref{superfields}) do not respect this symmetry. Hence, in the
``$\Phi-\Psi$" formalism, the $U(1)$ R-symmetry is broken to
$\mathbb{Z}_{2}$, which corresponds to the little group in
three-dimensions. This corresponds to multiplying bosons by +1 and
fermions by -1.

The $SO(2)$ R-symmetry of the theory is realized by the superamplitudes as follows. It is easy to see that the 4-pt. superamplitude is an eigenfunction of the R-symmetry generator
\[
R=\sum_{i=1}^{n}\left(\eta_{i}\frac{\partial}{\partial\eta_{i}}+\bar{\eta}_{i}\frac{\partial}{\partial\bar{\eta}_{i}}\right),
\]with eigenvalue 2. Hence, $R-2$  is a symmetry of the 4-pt. superamplitude. This corresponds to a $U(1)=SO(2)$ R-symmetry. We expect this symmetry to persist for higher point amplitudes. In particular, we expect that the $n$-pt. amplitude will be annihilated by $R-n/2$.

 \subsubsection{$\Phi-\Phi^\dagger$ Formalism}
 To make contact with the $\Phi-\Phi^\dagger$ formalism for four
 dimensional SYM theories, we first recall that in the notation of
 \cite{Elvang:2011fx} four dimensional $mhv$ amplitudes (with negative
 helicity particles in the $i$ and $j$ slots of the color ordered
 amplitude) correspond to \be \mathcal{A}^{mhv}_{i,j} = \langle
 ...\Phi^\dagger_i...\Phi^\dagger_j...\rangle \ee where the other
 entries correspond to $\Phi$.  For four point amplitudes \be
 \mathcal{A}^{mhv}_{i,j} = \tilde \Omega(i,j)\left(\tprods{i}{j} +
 \tprods{i}{k}\bar \eta_j \eta _k - \tprods{j}{k}\bar \eta _i \eta_k -
 \frac{1}{2}\tprods{k}{l}\bar \eta_i\bar
 \eta_k\eta_k\eta_l\right)\label{n=2massless} \ee where for the four
 dimensional theory -- as well as the massless three dimensional SYM
 theory -- the prefactor $\Omega $ is given by the famous Parke-Taylor
 relation: \be \tilde \Omega (i,j) =
 \frac{\tprods{i}{j}^3}{\tprods{1}{2}\tprods{2}{3}\tprods{3}{4}\tprods{4}{1}}.
 \ee Even though the massive $SCS$ theory is not known to be obtainable
 as a dimensional reduction of a higher dimensional gauge theory the
 massive supersymmetry algebra (\ref{n=2c}) can be regarded as a
 dimensional reduction of the four dimensional massless supersymmetry
 algebra (where the two $SU(2)$'s of the $d=4$ Lorentz group are
 identified and fourth components of all the physical momenta are
 fixed to be $m$). It is thus expected that the kinematic
 constraints relating the different components of the superamplitudes
 for the massive $CSM$ theory can be cast in a $\Phi-\Phi^\dagger$
 form as in $d=4$.  Indeed, after defining the adjoint superfield $
 \Phi^\dagger = a_- +\eta \chi_-$ in our notation, it is readily seen
 that \be \mathcal{A}^{CSM}_{i,j} = \Omega(i,j)\left(\tprods{\bar
   i}{\bar j} + \tprods{\bar i}{k} \eta_j \bar \eta _k - \tprods{\bar
   j}{k}\eta _i \bar \eta_k - \frac{1}{2}\tprods{ k}{
   l}\eta_i\eta_k\bar \eta_k\bar \eta_l\right)\label{n=2massive} \ee
 correctly reproduces all the relations between the massive amplitudes
 given above. The prefactor $\Omega $ can be read off once any of the
 known four-point component amplitudes are known. For our present
 purposes, they can be determined in terms of the four-fermion
 amplitudes computed in appendix \ref{4fermi}.

%*******************************************************************

\section{Yang-Mills-Chern-Simons amplitudes} \label{ymcsamp}

In this section, we will describe various three and four-point
tree-level color-ordered amplitudes of the $\mathcal{N}=2$ YMCS
theory. In particular, we compute all of the three and four-point
amplitudes without external gauge fields, and obtain the remaining
4-point amplitudes using the on-shell superalgebra
(\ref{undeformed})\footnote{Note that the on-shell superalgebra implies constraints on the 4-point
amplitudes but not on the 3-point amplitudes. This has to do with the
fact that the algebra is only valid when the external momenta are
real. In the case of three-point amplitudes one necessarily needs to
continue the amplitudes to complex momenta.}. In the end of this section, we describe the difficulties associated with computing on-shell YMCS amplitudes with external gauge fields using Feynman diagrams.  

%%%%%%%%%%%%%%%%%%%%%%%%%%%%%%%%%%%%%%%%%%%%%%%%%%%%%%%%%
\subsection{3-point amplitudes} \label{amp3}

The colour ordered 3-pt. amplitudes are defined for
completely general fields $\phi_{{\cal A}_i}$ by the expression
\bsp
\Bigl\la {\phi^{a_1}_{{\cal A}_1}}^\dag(p_1)\,& {\phi^{a_2}_{{\cal A}_2}}^\dag(p_2)\, 
{\phi^{a_3}_{{\cal A}_3}}^\dag(p_3)\Bigr\ra\\
& = 2ie \left\la\phi_{{\cal A}_1}\phi_{{\cal A}_2} \phi_{{\cal A}_3} \right\ra\Tr[T^{a_1}T^{a_2}T^{a_3}] + \ldots,
\end{split}
\ee
where the momenta are all in-going and ${\phi^{a}_{{\cal A}}}^\dag(p)$
is the creation operator for the associated field.

The only 3-pt. amplitude which does not have external gauge fields is
\bsp
&\la \Psi_{A_1} \Psi_{A_2} \,\Phi \ra = -\e_{A_1A_2}\, \bar
v(p_2) u(p_1) = 
-\e_{A_1A_2}\, \la 1 2  \ra.
\end{split}
\label{3pt}
\ee
Rearrangement of the fields is achieved using 
\bsp
&\left\la\phi_{{\cal A}_1}\phi_{{\cal A}_2} \phi_{{\cal A}_3} \right\ra
= -\la\phi_{{\cal A}_2}\phi_{{\cal A}_1} \phi_{{\cal A}_3} \ra\\
&\left\la\phi_{{\cal A}_1}\phi_{{\cal A}_2} \phi_{{\cal A}_3} \right\ra
= \la\phi_{{\cal A}_2}\phi_{{\cal A}_3} \phi_{{\cal A}_1} \ra.\\
\end{split}
\ee
The SUSY algebra does not help us determine the remaining 3-pt. amplitudes from (\ref{3pt}). 

Note that the amplitude in (\ref{3pt}) has $SO(2)$ R-symmetry which rotates the
two fermions. This symmetry follows from the $SO(2)$ R-symmetry in the
fermionic sector of the Lagrangian and should therefore hold for
higher-point amplitudes, as we will demonstrate at 4-point. The form of this amplitude will be useful for deducing whether or not the BCFW recursion relations are applicable to this massive gauge theory. We shall return to this issue later.

%%%%%%%%%%%%%%%%%%%%%%%%%%%%%%%%%%%%%%%%%%%%%%%%%%%%%%%%%%%%%%%%%%%%%%%%%%%
\subsection{4-point amplitudes}
In this section, we compute various tree-level 4-pt. amplitudes of the YMCS theory. One may determine the remaining amplitudes using the following
rearrangement rules
\bsp
&\la\phi_{\cal D}\, \phi_{\cal C} \,\phi_{\cal B}\,\phi_{\cal A}\ra = 
(-1)^{\text{f.e.}}\la\phi_{\cal A}\, \phi_{\cal B}\,
\phi_{\cal C} \,\phi_{\cal D}\ra
~\text{with}~ p_1 \lr p_4,~p_2 \lr p_3,\\
&\la \phi_{\cal B}\,
\phi_{\cal C} \,\phi_{\cal D}\, \phi_{\cal A} \ra 
= (-1)^{\text{f.e.}}\la \phi_{\cal A}\, \phi_{\cal B}\,
\phi_{\cal C} \,\phi_{\cal D}\ra ~\text{with}~ 
p_1 \to p_4,~p_2\to p_1,~ p_3\to p_2,~p_4 \to p_3,\\
&\la\phi_{\cal A}\, \phi_{\cal C}\, \phi_{\cal B} \,\phi_{\cal D} \ra =
  -(-1)^{\text{f.e.}}\la\phi_{\cal A}\, \phi_{\cal B}\, \phi_{\cal C}
  \,\phi_{\cal D} \ra ~\text{with}~ p_2 \lr p_3\\
&\qquad\qquad\qquad~~~
 - (-1)^{\text{f.e.}}\la\phi_{\cal A}\,\phi_{\cal C} \,\phi_{\cal D}
\,  \phi_{\cal B} \ra  ~\text{with}~ 
p_3 \lr p_4,
\end{split}
\ee 
where $\phi_{\cal A}$ indicates a general field and ``f.e.'' means the
number of times fermions (if present) are exchanged in the reordering.

We begin by computing the 4-fermion amplitudes. Then we compute two
fermion--two boson amplitudes, followed by 4-boson amplitudes.

\subsubsection{Four fermion amplitudes}

The calculation of the 4-fermion amplitudes of the YMCS theory is
described in appendix \ref{4fermi}. We obtain 
\bsp\label{YMCSff}
&\la\chi_+\chi_+\chi_-\chi_-\ra =\la\chi_-\chi_-\chi_+\chi_+\ra
=-\frac{2\la 3 4\ra}{u+m^2}\left[\la 1 2\ra+im\frac{\la 4 2\ra}{\la 4
    \bar 1\ra}\right],\\ &\la\chi_+\chi_-\chi_-\chi_+\ra
=\la\chi_-\chi_+\chi_+\chi_-\ra =\frac{2\la4 1\ra}{s+m^2}\left[\la 2
  3\ra+im\frac{\la 1 3\ra}{\la 1 \bar
    2\ra}\right],\\ &\la\chi_+\chi_-\chi_+\chi_-\ra
=\la\chi_-\chi_+\chi_-\chi_+\ra =\frac{2\la13\ra}{s+m^2}\left[\la 4
  2\ra-im\frac{\la14 \ra}{\la 1 \bar
    2\ra}\right]\\ &\qquad\qquad\qquad\qquad\qquad\qquad~~\,
-\frac{2\la42\ra}{u+m^2}\left[\la 3 1\ra-im\frac{\la43 \ra}{\la 4 \bar
    1\ra}\right],
\end{split}
\ee
where $s=(p_1+p_2)^2$, $t=(p_1+p_3)^2$, $u=(p_1+p_4)^2$.  

It is interesting to consider the massless limit of the four-fermion
amplitude. In the strict $m=0$ limit, we should recover the ${\cal
  N}=2$ SYM amplitude computed in eq. (3.20) of \cite{AADY1}, and
indeed that is what is found here. At the next order, ${\cal O}(m)$,
we find that the massive spinor products may not be expressed using
massless spinor products. Using the first amplitude above as an
example, we find that the massless limit is
\be\label{m0lim}
\la\chi_+\chi_+\chi_-\chi_-\ra =-2
\frac{\la12\ra^2}{\la23\ra\la41\ra} + {\cal O}(m),
\ee
where the spinor brackets are massless.

%---------------------------------------------------------%
\subsubsection{Two fermion -- two boson amplitudes}

We continue with the two fermion -- two boson amplitudes. In what
follows, the subscripts appearing on the spinors $u$ and $v$ refer to
particle (i.e. leg) number. Note that perturbation theory using the
mode expansions (\ref{ymcs-mode}) is consistent with the on-shell
algebra for amplitudes without external gauge fields and can therefore
be used to compute $\la \chi_+ \chi_- \Phi \Phi \ra$. In particular,
we obtain
\bsp
\la \chi_+ \chi_- \Phi \Phi \ra &=\la \chi_- \chi_+ \Phi \Phi \ra\\
& =
 \frac{\bar v_1 {\not p_4}\,
  u_2}{u+m^2}
-\frac{1}{s(s+m^2)}\Bigl(-2\,s\, \bar v_1 {\not p_4}\,u_2
+2im\,\e_{\m\n\r}\,p_4^\m\, p_3^\n\, \bar v_1 \g^\r\,u_2 \Bigr)\\
&=  -\frac{1}{2(u+m^2)} \bigl( \la 1 4\ra\la\bar 4 2\ra
+\la 1 \bar 4\ra \la 4  2\ra\bigr)\\
&- \frac{2(2m^2+s)}{s+m^2} \frac{\la 2 \bar 3\ra\la
  3\bar 1\ra}{\la \bar 1 \bar 2\ra\la \bar 1  2\ra} +\frac{im}{s+m^2}\frac{\la 1\bar
  3\ra\la\bar 1 3\ra-\la 2 \bar 3\ra\la\bar 2 3\ra}{\la \bar 1\bar 2 \ra}.
\end{split}
\ee

The two fermion -- two boson amplitudes with an external gauge field may be determined using the algebra
(\ref{undeformed}). Specifically one finds 
\bsp
&\la\Phi \chi_+ A \chi_-\ra = i\frac{\la\bar 4\bar 1\ra}{\la\bar 4\bar
  3\ra}\la\Psi_2\Psi_2\Psi_1\Psi_1\ra + i\frac{\la 2\bar 4\ra}{\la\bar
  4\bar 3\ra}\la\Phi\Phi\chi_+\chi_-\ra\\
&= i\frac{\la\bar 4\bar 1\ra}{\la\bar 4\bar
  3\ra}\Biggl[
\frac{\la 41\ra\la 23\ra}{u+m^2}+\frac{1}{s+m^2}\Biggl(
2\la  2 3\ra\la 4 1\ra-
\la  1 2\ra\la 3 4\ra
-2im\,\frac{\la 1 3\ra\la 14\ra}{\la1\bar 2\ra}\Biggr)\Biggr]\\
&+ i\frac{\la 2\bar 4\ra}{\la\bar
  4\bar 3\ra}\Biggl[
 -\frac{1}{2(u+m^2)} \bigl( \la 3 2\ra\la\bar 2 4\ra
+\la 3 \bar 2\ra \la 2  4\ra\bigr)\\
&\qquad\qquad
- \frac{2(2m^2+s)}{s+m^2} \frac{\la 4 \bar 1\ra\la
  1\bar 3\ra}{\la \bar 3 \bar 4\ra\la \bar 3  4\ra} +\frac{im}{s+m^2}\frac{\la 3\bar
  1\ra\la\bar 3 1\ra-\la 4 \bar 1\ra\la\bar 4 1\ra}{\la \bar 3\bar 4
  \ra}
\Biggr].
\end{split}
\ee
\bsp\label{pAAm}
&\la\chi_+ AA \chi_-\ra =
-\frac{\la 4 1\ra\la\bar 4 \bar 1\ra}{\la\bar 2 4\ra\la\bar 4\bar
  3\ra} \la \Psi_2\Psi_2\Psi_1\Psi_1\ra +\frac{\la 4 3\ra}{\la\bar 2
  4\ra}\la\Psi_1\Psi_2\Psi_2\Psi_1\ra
-\frac{\la 4 1\ra\la 2 \bar 4\ra}{\la \bar 2  4\ra\la\bar 4\bar
  3\ra} \la \Phi\Phi \chi_+\chi_-\ra\\
&=\frac{1}{\la\bar 2\bar 1\ra} \Biggl[
-2\frac{\la 41\ra\la 23\ra}{\la\bar 3 1\ra}(s+2m^2) +2\frac{\la 12
  \ra\la 34\ra}{\la\bar 3 1\ra}(s+4m^2)\\
&\qquad\qquad
-im\left(\la 32\ra\la 34\ra-2\frac{\la 42\ra\la 34\ra}{\la\bar 3
  1\ra\la 4\bar 1\ra}(s+4m^2)\right)\Biggr]\frac{1}{u+m^2}\\
&+\frac{1}{\la\bar 4\bar 3\ra}\Biggl[
\frac{\la 12\ra\la 34\ra}{\la\bar 2 4\ra}(s-u)
-2\frac{\la 23\ra\la 41\ra}{\la\bar 2 4\ra}(t+s)
-2\frac{\la 23\ra\la 4 \bar 1\ra\la 1 \bar 3\ra}{\la\bar 1\bar 2\ra\la
  \bar 3 4\ra}(s+2m^2)\\
&\qquad\qquad
+2im\frac{\la 13\ra\la 14\ra}{\la\bar 2 4\ra\la 1 \bar 2\ra}(t+s)
+im\frac{\la 23\ra}{\la\bar 1\bar 2\ra}(u-t)\Biggr]\frac{1}{s+m^2}.
%&=-\frac{\la 4 1\ra\la\bar 4 \bar 1\ra}{\la\bar 2 4\ra\la\bar 4\bar
%  3\ra} \Biggl[
%\frac{\la 41\ra\la 23\ra}{u+m^2}+\frac{1}{s+m^2}\Biggl(
%2\la  2 3\ra\la 4 1\ra-
%\la  1 2\ra\la 3 4\ra
%-2im\,\frac{\la 1 3\ra\la 14\ra}{\la1\bar 2\ra}\Biggr)\Biggr]\\
%&-\frac{\la 4 3\ra}{\la\bar 2
%  4\ra}\Biggl[
%\frac{\la 12\ra\la 34\ra}{s+m^2}+\frac{1}{u+m^2}\Biggl(
%2\la  1 2\ra\la 3 4\ra-
%\la  4 1\ra\la 2 3\ra
%-2im\,\frac{\la 4 2\ra\la 43\ra}{\la4\bar 1\ra}\Biggr)\Biggr]\\
%&-\frac{\la 4 1\ra\la 2 \bar 4\ra}{\la \bar 2  4\ra\la\bar 4\bar
%  3\ra}\Biggl[
% -\frac{1}{2(u+m^2)} \bigl( \la 3 2\ra\la\bar 2 4\ra
%+\la 3 \bar 2\ra \la 2  4\ra\bigr)\\
%&\qquad\qquad
%- \frac{2(2m^2+s)}{s+m^2} \frac{\la 4 \bar 1\ra\la
%  1\bar 3\ra}{\la \bar 3 \bar 4\ra\la \bar 3  4\ra} +\frac{im}{s+m^2}\frac{\la 3\bar
%  1\ra\la\bar 3 1\ra-\la 4 \bar 1\ra\la\bar 4 1\ra}{\la \bar 3\bar 4
%  \ra}
%\Biggr].
\end{split}
\ee

%---------------------------------------------------------%
\subsubsection{Four boson amplitudes}

The four $\Phi$ amplitude may be computed using perturbation theory
and one finds
\bsp
\la \Phi \Phi \Phi \Phi \ra &= 
\frac{(t-u)s-4im\e_{\m\n\r}p_1^\m p_2^\n p_3^\r}{s(s+m^2)}
+\frac{(t-s)u+4im\e_{\m\n\r}p_1^\m p_2^\n p_3^\r}{u(u+m^2)}\\
&=\frac{\la 1\bar 4\ra\la \bar 1 4 \ra -  \la1\bar 3\ra\la \bar 1 3 \ra}{s+m^2}  +\frac{2im\la 1
\bar 2\ra\la 2\bar 3\ra\la 3 \bar 1\ra}{s(s+m^2)}\\
&+\frac{\la 1\bar 2\ra\la \bar 1 2 \ra -  \la1\bar 3\ra\la \bar 1 3\ra}{u+m^2}  -\frac{2im\la 1
\bar 2\ra\la 2\bar 3\ra\la 3 \bar 1\ra}{u(u+m^2)}.
\end{split}
\ee
The four boson amplitudes with external gauge fields
may be gotten using the algebra in (\ref{undeformed}). One finds
\bsp
&\la\Phi\Phi A A\ra = -\frac{\la\bar 1\bar 2\ra}{\la\bar 3\bar 4\ra}
\la \Psi_2\Psi_2\Psi_1\Psi_1\ra\\
&= -\frac{\la\bar 1\bar 2\ra}{\la\bar 3\bar 4\ra}
\Biggl[
\frac{\la 41\ra\la 23\ra}{u+m^2}+\frac{1}{s+m^2}\Biggl(
2\la  2 3\ra\la 4 1\ra-
\la  1 2\ra\la 3 4\ra
-2im\,\frac{\la 1 3\ra\la 14\ra}{\la1\bar 2\ra}\Biggr)\Biggr].
\end{split}
\ee
\bsp
\la A A A A\ra &= \frac{\la 3 2\ra}{\la\bar 1 3\ra}\la\chi_+\chi_- A
A\ra +\frac{\la 3 4\ra}{\la\bar 1 3 \ra} \la \chi_+ A A\chi_-\ra\\
&=-\frac{\la 3 2\ra}{\la\bar 1 3\ra}\la\chi_+A A
\chi_-\ra_{i\to i+1} +\frac{\la 3 4\ra}{\la\bar 1 3 \ra} \la \chi_+ A A\chi_-\ra,
\end{split}
\ee
where $i\to i+1$ indicates momentum relabelling, and $\la \chi_+ A
A\chi_-\ra$ is given in (\ref{pAAm}).

We end this section by pointing out that the computation of amplitudes
involving external gauge fields using Feynman diagrams is
significantly more complicated than the other computations presented
in this paper. The complications have to do with defining a mode
expansion for the gauge fields that is compatible with both the
non-commutativity of the spatial components of the vector potential as
well as the Gauss law constraints mentioned in section
\ref{ymcsformal}. In particular, it has been argued in
\cite{Haller1,Haller2} that the canonical commutation relations of the
gauge field cannot be satisfied if the mode expansion of the gauge
field only contains the modes of an on-shell massive scalar
field. Auxiliary fields must also appear in the mode expansion in
order for the theory to be consistently quantized and for the
non-commutativity of the gauge fields to be respected, making the use
of Feynman diagrams extremely unweildy. We are able to circumvent this
difficulty in the results presented above by using the on-shell SUSY
algebra -- the algebra was shown to be consistent with both the
off-shell superalgebra as well as the canonical quantization procedure
in section \ref{ymcsformal} -- to determine the amplitudes
containing the external gauge fields.

It would be extremely desirable to have a spinor helicity framework for the computations of gauge field amplitudes in YMCS theories (with and without supersymmetry) using Feynman diagrams efficiently. We hope to analyze this issue in further detail  elsewhere.

%There are various difficulties that arise when trying to compute YMCS amplitudes with external gauge fields using Feynman diagrams. %These complications stem from the difficulty in defining a mode expansion for the gauge field. For instance, one might try expanding the %gauge field as
%
%\be
%A^a_\mu(x) = \int
%\frac{d^2p}{(2\pi)^2}\frac{1}{\sqrt{2p^0}}\left(\epsilon_{\mu}(p)a_1^{a\dagger}(p)e^{ip\cdot x}
%+ \epsilon^*_{\mu}(p) a^a_1(p)e^{-ip\cdot x} \right),
%\ee
%with the polarization vector given by
%\be
%\epsilon^{\mu}(p)=\frac{\bar{u}(p)\gamma^{\mu}v(p)}{2\sqrt{2}m}.
%\ee
%

%***************************************************************************%

%%%%%%%%%%%%%%%%%%%%%%%%%%%%%%%%%%%%%%%%%%%%%%%%%%%%%%%%%%%%%%%%%%

\section{BCFW for mass-deformed three-dimensional theories} \label{bcfwsection}

A very useful tool for computing scattering amplitudes are the BCFW recursion relations, which allow one to construct higher point on-shell amplitudes from lower-point on-shell amplitudes \cite{Britto:2005fq}. The BCFW recursion relations in $d \geq 4$ do not hold in 3d, however, even in the mass-deformed case. In $d \geq 4$ one derives the recursion relations by deforming two external legs of an on-shell amplitude as follows:
\[ 
p_{i}\rightarrow p_{i}+zq,\,\,\, p_{j}\rightarrow p_{j}-zq
\]
where $z$ is a complex number and $q$ is some vector. In order for the momenta to remain on-shell for general $z$, we must impose the following conditions on $q$:
\[
q\cdot p_{i}=q\cdot p_{j}=q^{2}=0.
\]
In 3d, the only solution is $q=0$. Hence, the usual BCFW deformation does not apply in 3d, even in the mass-deformed case. In order to define a two-line deformation, we must allow the deformation to be non-linear. The BCFW recursion relations for massless 3d theories were derived in \cite{Gang:2010gy}. In this section, we will propose BCFW recursion relations for massive 3d theories. 

\subsection{Two-line deformation}

The BCFW recursion relations follow from deforming the momenta of
two external legs of an on-shell amplitude. Suppose we deform legs $i$ and $j$. The deformation
must preserve the total momentum 

\begin{equation}
\left(p_{i}+p_{j}\right)^{\alpha\beta}=\lambda_{i}^{(\alpha}\bar{\lambda}_{i}^{\beta)}+\lambda_{j}^{(\alpha}\bar{\lambda_{j}}^{\beta)}{\normalcolor .}\label{eq:momentum}\end{equation}
The deformation must also preserve the following
two conditions 
\begin{equation}
\left\langle i\bar{i}\right\rangle ^{2}=-4m^{2},\,\,\,\left\langle j\bar{j}\right\rangle ^{2}=-4m^{2}.\label{eq:massshell}
\end{equation}
We will assume that all external particles of an on-shell amplitude have the same mass. 

If the external particles are massless, then the momentum is given
by 
\[
\left(p_{i}+p_{j}\right)^{\alpha\beta}=\lambda_{i}^{\alpha}\lambda_{i}^{\beta}+\lambda_{j}^{\alpha}\lambda_{j}^{\beta}.
\]
In this case, the BCFW deformation is given by \cite{Gang:2010gy}
\begin{equation}
\left(\begin{array}{c}
\lambda_{i}\\
\lambda_{j}
\end{array}\right)\rightarrow\left(\begin{array}{cc}
\frac{1}{2}\left(z+z^{-1}\right) & \frac{i}{2}\left(z-z^{-1}\right)\\
-\frac{i}{2}\left(z-z^{-1}\right) & \frac{1}{2}\left(z+z^{-1}\right)
\end{array}\right)\left(\begin{array}{c}
\lambda_{i}\\
\lambda_{j}
\end{array}\right),
\label{deform1}
\end{equation}
where $z$ is an arbitrary complex number. The deformation above clearly conserves
momentum since it is an orthogonal transformation. For the mass deformed
case, there is a natural generalization. We simply deform the antiholomorphic spinors in the same way as the holomorphic ones in (\ref{deform1})   
\begin{equation}
\left(\begin{array}{c}
\bar{\lambda}_{i}\\
\bar{\lambda}_{j}
\end{array}\right)\rightarrow\left(\begin{array}{cc}
\frac{1}{2}\left(z+z^{-1}\right) & \frac{i}{2}\left(z-z^{-1}\right)\\
-\frac{i}{2}\left(z-z^{-1}\right) & \frac{1}{2}\left(z+z^{-1}\right)
\end{array}\right)\left(\begin{array}{c}
\bar{\lambda}_{i}\\
\bar{\lambda}_{j}
\end{array}\right).
\label{deform2}
\end{equation}

It is easy to see that (\ref{deform1}) and (\ref{deform2}) preserve momentum in (\ref{eq:momentum}). Furthermore, after these transformations we see
that
\[
\left\langle i\bar{i}\right\rangle \rightarrow\left\langle i\bar{i}\right\rangle +\left(\left\langle i\bar{j}\right\rangle -\left\langle \bar{i}j\right\rangle \right)\frac{i}{4}(z^2-z^{-2}),\,\,\,\left\langle j\bar{j}\right\rangle \rightarrow\left\langle j\bar{j}\right\rangle -\left(\left\langle i\bar{j}\right\rangle -\left\langle \bar{i}j\right\rangle \right)\frac{i}{4}(z^2-z^{-2}).
\]
Note that 
\[
\left\langle i\bar{j}\right\rangle =e^{i\kappa}\sqrt{\left(p_{i}+p_{j}\right)^{2}},\,\,\,\left\langle \bar{i}j\right\rangle =e^{-i\kappa}\sqrt{\left(p_{i}+p_{j}\right)^{2}},
\]
where $e^{i\kappa}$ is some $U(1)$ phase. Also note that we can we can
redefine $\lambda_i$ and $\bar{\lambda}_i$ by a phase since
$p_i^{\alpha\beta}=\lambda^{(\alpha}_i\bar{\lambda}^{\beta)}_i$ is
invariant under
$\left(\lambda_i,\bar{\lambda}_i\right)\rightarrow\left(e^{i\omega}\lambda_i,e^{-i\omega}\bar{\lambda}_i\right)$.
Hence, by taking
$\left(\lambda_{i},\bar{\lambda}_{i}\right)\rightarrow\left(e^{-i\kappa}\lambda_{i},e^{i\kappa}\bar{\lambda}_{i}\right)$,
this will set $\left\langle i\bar{j}\right\rangle =\left\langle
\bar{i}j\right\rangle $ and the mass-shell conditions in
(\ref{eq:massshell}) will be preserved. After fixing the phases of the $(\lambda,\eta)$ variables, there is still a residual $U(1)$ symmetry which rotates all the $(\lambda,\eta)$ variables in the same way. In the mass-deformed Chern-Simons theory, this $U(1)$ phase is then fixed once we define the superfields, as explained in section \ref{superamp}. Note that the deformations in
(\ref{deform1}) and (\ref{deform2}) also preserve $\left\langle
ij\right\rangle$ and $\left\langle \bar{i}\bar{j} \right\rangle$.

To generalize this to a super-BCFW shift, consider the definition
of supermomentum in (\ref{supermomentum})\[
q=\lambda\bar{\eta}+\bar{\lambda}\eta.\]
Then the sum of the supermomenta of the particles which are being shifted
is given by
\[
q_{i}+q_{j}=\lambda_{i}\bar{\eta}_{i}+\bar{\lambda}_{i}\eta_{i}+\lambda_{j}\bar{\eta}_{j}+\bar{\lambda}_{j}\eta_{j}.\]
The supermomentum will be preserved if we apply the same BCFW deformation
to the fermionic coordinates as we do to the bosonic coordinates of
the on-shell superspace 
\[
\left(\begin{array}{c}
\eta_{i}\\
\eta_{j}\end{array}\right)\rightarrow\left(\begin{array}{cc}
\frac{1}{2}\left(z+z^{-1}\right) &  \frac{i}{2}\left(z-z^{-1}\right)\\
- \frac{i}{2}\left(z-z^{-1}\right) & \frac{1}{2}\left(z+z^{-1}\right)\end{array}\right)\left(\begin{array}{c}
\eta_{i}\\
\eta_{j}\end{array}\right),
\]
\begin{equation}
\left(\begin{array}{c}
\bar{\eta}_{i}\\
\bar{\eta}_{j}\end{array}\right)\rightarrow\left(\begin{array}{cc}
\frac{1}{2}\left(z+z^{-1}\right) &  \frac{i}{2}\left(z-z^{-1}\right)\\
- \frac{i}{2}\left(z-z^{-1}\right) & \frac{1}{2}\left(z+z^{-1}\right)\end{array}\right)\left(\begin{array}{c}
\bar{\eta}_{i}\\
\bar{\eta}_{j}\end{array}\right).
\label{etadef}
\end{equation}

%%%%%%%%%%%%%%%%%%%%%%%%%%%%%%%%%%%%%%%%%%%%%%%%%%
\subsection{Recursion relation}

After performing the BCFW deformation, the amplitude becomes a function of $z$.
Assuming the amplitude vanishes when $z\rightarrow\infty$, we have
the following
\begin{equation}
\oint_{|z|=\infty}\frac{A(z)d z}{z-1}=0{\normalcolor .}\label{contour-int}
\end{equation}
On the other
hand, this contour integral must also be equal to the sum of the residues
of the integrand in the complex plane, which occur at $z=1$ and the
poles of $A(z)$. Near its poles, $A(z)$
factorizes into two on-shell amplitudes (denoted $A_{L}$ and $A_{R}$)
multiplied by a propagator. Hence, we find that
\begin{equation}
A(z=1)=-\frac{1}{2\pi i}\sum_{f,j}\int
d\eta\oint_{z_{f,j}}\frac{A_{L}(z,\eta)A_{R}(z,i\eta)}{\hat{p}_{f}(z)^{2}+m^2}\frac{1}{z-1},\label{eq:bcfw}\end{equation}
where the factorization channels are labeled by $f$, and $z_{f,j}$
corresponds to the $j$-th root of $\hat{p}_{f}(z)^{2}+m^2$. In
obtaining this formula, we assumed that all the external legs of the
on-shell scattering amplitudes have the same mass, $m$. The integral
$\int d\eta$ takes into account all the fields in the supermultiplet
which can appear in the propagator. Note that $A(z=1)$ corresponds to
the undeformed on-shell amplitude. Using (\ref{eq:bcfw}), we can
compute higher-point on-shell amplitudes from lower-point on-shell
amplitudes.

From the deformation in (\ref{deform1}) and (\ref{deform2}), one can see that in any
channel, $\hat{p}_{f}(z)^{2}+m^2$ has the following form

\[
\hat{p}_{f}(z)^{2}+m^2=a_{f}z^{-2}+b_{f}+c_{f}z^{2}{\normalcolor .}\]
Hence the roots are obtained by solving a quadratic equation in
$z^{2}$, see appendix \ref{sec:fact}.

%%%%%%%%%%%%%%%%%%%%%%%%%%%%%%%%%%%%%%%%%%%%%%%%%%%%%%%%%%%%%%%%%
\subsection{Large-$z$ behavior} \label{bcfwcheck}
In order for the recursion relation described in the previous section
to be applicable, the on-shell amplitudes must vanish after performing
the BCFW deformations in (\ref{deform1}), (\ref{deform2}), and
(\ref{etadef}) and taking the deformation parameter $z$ to
infinity.

The amplitudes of the YMCS theory do not generally have good large-$z$
behavior. Furthermore, it does not appear to be possible to combine
them into superamplitudes (which could in principle have better
large-$z$ behavior). Hence, our proposed BCFW recursion relation does
not appear to be applicable to the $\mathcal{N}=2$ YMCS theory. The
situation may improve for YMCS theories with more supersymmetry
however.

Although the 4-pt. component amplitudes of the CSM theory also do not
generally have good large-$z$ behavior, our proposed recursion relation may be applicable to the
superamplitudes of the CSM theory. In particular, the first 4-pt. superamplitude in
(\ref{eq:cssuperamp}) is $\mathcal{O}(1/z)$ when legs $(1,3)$ or $(2,4)$ are
shifted. In order to test this, one should
use the recursion relation to compute a 6-pt. superamplitude of the
CSM theory, and match various components of the superamplitude with
Feynman diagram calculations.

%%%%%%%%%%%%%%%%%%%%%%%%%%%%%%%%%%%%%%%%%%%%%%%%%%%%%%%%%%%%%%%%%%%
\subsection{Factorization of YMCS amplitudes from BCFW shift}

It is interesting to investigate the four-point amplitudes of the YMCS
theory in the vicinity of their poles, and to look for simple
factorization into two three-point amplitudes. As an illustrative
example, we will look at one of the four-fermion amplitudes
\be\label{amp}
{\cal A}(1) = \la \chi_+ \chi_+ \chi_- \chi_- \ra = \frac{2\la 43 \ra}{s_{23}+m^2}
\left[ \la 12 \ra + im \frac{\la 42 \ra}{\la 4 \bar 1 \ra}\right],
\ee
and perform the BCFW shift on legs 1 and 2. We find
\bsp
{\cal A}(z) =&\frac{2z^2}{m^2(s_-+\sqrt{s_-^2-s_+^2})}\\
&\times\frac{2\la 43 \ra}{(z^2-z_1^2)(z^2-z_2^2)} \left[
\la 12 \ra + im \frac{(z^2+1)\la 42 \ra-i(z^2-1)\la 41 \ra  }
{(z^2+1)\la 4 \bar 1 \ra + i(z^2-1)\la 4 \bar 2\ra}\right],
\end{split}
\ee
where (see appendix \ref{sec:fact} for details)
\be
s_\pm = \frac{1}{2m^2}\left(s_{13} \pm s_{23}\right),
\ee
and where the massive poles corresponding to $s_{23}=-m^2$ are found at
$z=\pm z_1$ and $z=\pm z_2$ where
\be\label{rootss}
\{z_1^2,z_2^2\} = \left\{\frac{1}{2}\frac{\left(\sqrt{2s_++1}+ 1\right)^2}
{s_-+\sqrt{s_-^2-s_+^2}},\,
\frac{1}{2}\frac{\left(\sqrt{2s_++1}- 1\right)^2}
{s_-+\sqrt{s_-^2-s_+^2}}
\right\}.
\ee
Note that the massless pole coresponding to $s_{23}=0$ has a vanishing
residue\footnote{This can be seen by noting that the $\la 4 \bar 1
  \ra$ appearing in the second term in (\ref{amp}) is proportional to
  $\sqrt{s_{23}}$.}; this is consistent with the fact that the YMCS
gauge field has only a single, massive degree of freedom. 

In order to understand the factorization we should associate the
residues at the massive poles with the product of ``left'' and ``right''
three-point amplitudes ${\cal A}_L$ and ${\cal A}_R$. We should be able
to see both a contribution from two fermion-fermion-scalar amplitudes
and also one from two fermion-fermion-gauge field amplitudes. We
therefore write the deformed amplitude in the following way
\bsp\label{thingy}
{\cal A}(z) = 
&\frac{2z^2}{m^2(s_-+\sqrt{s_-^2-s_+^2})}
\frac{1}{(z^2-z_1^2)(z^2-z_2^2)}\Bigl[
 \Bigl(\la 1' 4 \ra \la 2' 3 \ra\Bigr) \\
&+\Bigl(-2\la 12 \ra \la 34\ra-\la 1' 4 \ra \la 2' 3
  \ra+2im\la 42' \ra\la 43\ra/\la  4 \bar 1' \ra\Bigr)\Bigl],
\end{split}
\ee
where the prime denotes the BCFW rotated spinor. The first term
(enclosed in rounded parentheses) is the scalar exchange
and the follwing factorization
\be
{\cal A}_L(z_1) = \la 1' 4\ra,\quad
{\cal A}_R(z_1) = \la 2' 3\ra,
\ee
matches with the three-point amplitudes calculated for
fermion-fermion-scalar scattering in section \ref{amp3}.

The second rounded-parentheses term in (\ref{thingy}) corresponds to
the gauge field exchange and so obviously the product of the two
fermion-fermion-gauge field three-point functions yield
\be
{\cal A}_L {\cal A}_R = -2\la 12 \ra \la 34\ra-\la 1' 4 \ra \la 2' 3
  \ra+2im\la 42' \ra\la 43\ra/\la  4 \bar 1' \ra.
\ee
There is some freedom in how to factorize this expression into left
and right components -- computing the relevant three-point functions
using the techniques of \cite{Haller1,Haller2} would allow one to determine this
factorization, and we leave this issue as further work.

%%%%%%%%%%%%%%%%%%%%%%%%%%%%%%%%%%%%%%%%%%%%%%%%%
\section{Conclusion} \label{conclusion}

In this paper, we study scattering amplitudes of mass-deformed
three-dimensional gauge theories. In particular, we focus on
mass-deformed Chern-Simons and Yang-Mills-Chern-Simons theories with
$\mathcal{N}=2$ supersymmetry. Note that the mass deformations in
these theories preserve locality, Lorentz invariance, and gauge
invariance. We derive the superalgebras which underlie the scattering
matrices of the $\mathcal{N}=2$ mass-deformed CSM theory and YMCS
theory and show that the on-shell supersymmetry algebras for the two
theories are fundamentally different. In particular, the algebra for
YMCS contains no mass-deformation.

Using perturbative techniques and on-shell superalgebras, we compute 3
and 4-pt. tree-level colour-ordered amplitudes in these theories (note
that the odd point amplitudes of the CSM theory vanish). For the CS
theory, we find that perturbation theory gives results that are
consistent with the mass-deformed on-shell superalgebra. Further, we
find that the 4-pt. amplitudes of the CSM theory can be encoded in
very simple superamplitudes. On the other hand, for the YMCS theory we
are able to deduce all the four point amplitudes using a combination
of perturbative techniques and algebraic relations. Namely, we compute
all the amplitudes without external gluons perturbatively (and show
that they are consistent with the on-shell algebra) and deduce the
remaining 4-pt. amplitudes using the on-shell superalgebra in
(\ref{undeformed}).

We also propose a BCFW recursion relation for mass-deformed
three-dimensional gauge theories which reduces to the BCFW recursion
relation proposed in \cite{Gang:2010gy} in the massless limit. This
recursion relation involves deforming the supermomenta of two external
legs of an on-shell amplitude by a complex parameter $z$ and is only
applicable if the amplitude vanishes as $z\rightarrow\infty$. Although
the component amplitudes of the $\mathcal{N}=2$ CSM and YMCS theories
do not generally have good large-$z$ behavior, we find that one of the
4-pt. superamplitudes of the CSM theory exhibits good large-$z$
behavior, which suggests that the recursion relation may be applicable
to this theory.

There are a number of open questions that would be interesting to
address. First of all, it would be very desirable to understand how to
compute amplitudes with external gauge fields in the YMCS theory using
Feynman diagrams. In particular, it would be desirable to use Feynman
diagrams to compute the 3-pt. amplitudes with external gauge fields
and confirm the 4-pt. amplitudes with external gauge fields which we
deduced using the on-shell superalgebra. It would also be interesting
to test our BCFW proposal by using it to compute a
6-pt. superamplitude of the CSM theory and then compare it to a
Feynman diagram calculation.

Another interesting direction would be to extend our analysis to loop
amplitudes. Note that IR divergences of loop amplitudes are more
severe in three-dimensions than in four. On the other hand, we expect
that mass-deformations will lead to better IR behavior. It would also
be interesting to extend our analysis to mass-deformed theories with
more supersymmetry, like the mass-deformed ABJM theory, which has
$\mathcal{N}=6$ supersymmetry. If the amplitudes of YMCS theories with
$\mathcal{N}>2$ supersymmetry can be encoded in superamplitudes, then
the BCFW recursion relation proposed in this paper may be applicable
to these theories since superamplitudes generally have better
large-$z$ behavior than component amplitudes.

The techniques developed in this paper may also be useful for studying
the scattering amplitudes of three-dimensional gauge theories with
spontaneously broken gauge symmetry. In this case, masses are acquired
via the Higgs mechanism. In particular, it would be interesting to
study scattering amplitudes in the Coulomb branch of the ABJM theory
and see if they can be related to the amplitudes of maximal
three-dimensional SYM theory in some limit. There is already some
evidence that the amplitudes of three-dimensional SYM and ABJM
theory can be related order by order in perturbation theory in a
certain limit \cite{Lipstein:2012kd,Bianchi:2012ez}.

\section*{Acknowledgments}

We are grateful to V.P. Nair for useful discussions. AL is supported
by a Simons Postdoctoral Fellowship.

%%%%%%%%%%%%%%%%%%%%%%%%%%%%%%%%%%%%%%%%%%%%%%%%%%%%%%%%%%%%%%%%%%%%%%%%%%%%%%%%%%
\appendix

\section{Conventions, propagators and Feynman rules} \label{conventions}
\label{sec:app}
We work in $(-++)$ signature and use the following gamma matrices:
\be
\gamma ^\mu = \{i\sigma ^2, \sigma ^1, \sigma ^3\}.
\ee
The $SU(N)$ generators $t^a$ obey the following relations
\bsp
&t^a t^a = \frac{N^2-1}{2N} {\bf 1}, \quad \tr(t^at^b) = \frac{1}{2}
\d^{ab}, \quad [t^a,t^b] = i f^{abc} t^c,\\& f^{abc}f^{abd} =
N\d^{cd}
,\quad\{t^a,t^b\} = \frac{1}{N}\d^{ab} {\bf 1} + d^{abc} t^c.
\end{split}
\ee
The scalar and fermionic fields in this paper have mode expansions given by
\bsp\label{ymcs-mode}
& \Psi_{\a}(x)= \int
\frac{d^2p}{(2\pi)^2}\frac{1}{\sqrt{2p^0}}\left(v_\a(p) b^{\dagger}(p)e^{ip\cdot x} +
u_\a(p)b(p)e^{-ip\cdot x} \right),\\
&\Phi(x) = \int
\frac{d^2p}{(2\pi)^2}\frac{1}{\sqrt{2p^0}}\left(a_2^{\dagger}(p)e^{ip\cdot x}
+ a_2(p)e^{-ip\cdot x} \right),
\end{split}
\ee
where 
\bsp\label{spinors}
&v(p) = \frac{1}{\sqrt{p_0 -
p_1}}\matr{c}{p_2 + im\\ p_1 - p_0},\hspace{.5cm} u(p) =
\frac{1}{\sqrt{p_0 - p_1}}\matr{c}{p_2 -im\\ p_1 - p_0},\\
\end{split}
\ee
and we have neglected color and R-symmetry indices. There are many useful formulae involving spinors and gamma matrices
\bsp
&\gamma ^\mu \gamma ^\nu = \eta ^{\mu \nu} + \epsilon^{\mu \nu \rho}\gamma_\rho,\quad
\e^{\m\n\r}\e_{\g\d\r} = -\d^\m_\g \d^\n_\d + \d^\m_\d \d^\n_\g,\\
&\left(\g^\m\right)_{\a\b} \left(\g_\m\right)_{\g\d} = 2\d_{\a\d}\d_{\b\g}-\d_{\a\b}\d_{\g\d},\\
&\e^{\r\m\n} \left(\g_\m\right)_{\s\t}\left(\g_\n\right)_{\a\d}=2\left(\g^\r\right)_{\a\t}\d_{\s\d}-
\left(\g^\r\right)_{\a\d}\d_{\s\t} -
\left(\g^\r\right)_{\s\t}\d_{\a\d},\\
&u^*(p) = v(p),\quad \bar v = u^T\,C,\quad \bar u = v^T\,C,\\
&\bar v(p) v(p) = 2im,\qquad \bar u(p) u(p) = -2im,\quad \bar v(p) u(p) = 0 
= \bar u(p) v(p),\\
&{\not p}\, v = im v,\quad \bar v {\not p} = im \bar v,\quad
{\not p} \,u = -im u,\quad \bar u {\not p} = -im \bar u,\\
&\bar v(k) \g^\m u(p) = \bar v(k) u(p) \frac{im(p-k)^\m +
  \e^{\m\n\r}p_\n k_\r}{m^2+p\cdot k},\\
&\bar u(k) \g^\m u(p) = \bar u(k) u(p) \frac{im(p+k)^\m -
  \e^{\m\n\r}p_\n k_\r}{m^2-p\cdot k},\\
&\left| \bar u(p) u(k) \right|^2 = \left| \bar v(p) v(k) \right|^2 =
-(p+k)^2,\\
& \left| \bar u(p) v(k) \right|^2 = \left| \bar v(p) u(k) \right|^2 =
(p-k)^2,\\
&\bar v(p_i) v(p_j) = \la \bar j i \ra,\quad\bar u(p_i) u(p_j) = \la  j \bar i \ra,\\
&\bar u(p_i) v(p_j) = \la \bar i \bar j \ra,\quad\bar v(p_i) u(p_j) =
\la ij \ra,\\
&-\sqrt{-\frac{st}{u}}=2im-\frac{\left\langle 13\right\rangle \left\langle \bar{1}\bar{2}\right\rangle }{\left\langle \bar{2}3\right\rangle }=-\frac{\left\langle \bar{1}3\right\rangle \left\langle 1\bar{2}\right\rangle }{\left\langle \bar{2}3\right\rangle }.
\end{split}
\ee
For the $\mathcal{N}=2$ YMCS theory, the propagators are given by the following expressions
\bsp\label{props}
&\la A_\m^a (p) A_\n^b(-p)\ra  =-ie^2\d^{ab}\D_{\m\n}(p)= \frac{-ie^2\d^{ab}}{p^2(p^2+m^2)}\left(
p^2 \eta_{\m\n}-p_\m p_\n+im\e_{\m\n\r}p^\r\right),\\
&\la \Phi^a(p) \Phi^b(-p)\ra = \frac{-ie^2\d^{ab}}{p^2+m^2},\\
&\la \Psi_{A\a}^a (p) \Psi_{B\b}^b(-p)\ra = \frac{-ie^2\d^{ab}\d_{AB}}{p^2+m^2}\left[\left({\not
    p}+im\right)C^{-1}\right]_{\a\b},
\end{split}
\ee
where $C=\g^0$ is the charge conjugation matrix. In obtaining the
gauge field propagator, we have chosen Landau gauge. For the ${\cal
  N}=2$ massive Chern-Simons theory, the scalar and fermion propagators
are given by similar expressions and the gauge field propagator may be
read-off from the $m\to\infty$ limit of the YMCS gauge field
propagator
\be
\la A_\m^a (p) A_\n^b(-p)\ra_{CS}  = \frac{1}{\kappa}\frac{\e_{\m\n\r}\,p^\r}{p^2}.
\ee
Finally, we have made use of the following Feynman rules, where all momenta are in-going unless
explicitly indicated via an arrow, and where gluons, fermions, and
scalars are represented by wiggly, dashed, and solid lines
respectively
\vspace{0.5cm}
\[
\parbox{20mm}{
\begin{fmfgraph*}(50,50)
\fmftop{v1}
\fmfbottom{v2,v3}
%\fmffreeze
\fmf{scalar,label=$p_1$}{v1,vc1}
\fmf{scalar,label=$p_2$,label.side=left}{vc1,v2}
\fmf{photon,label=$p_3$,label.side=right}{v3,vc1}
\fmflabel{$a,A,\a$}{v1}
\fmflabel{$b,B,\b$}{v2}
\fmflabel{$c,\m$}{v3}
%\fmffreeze
%\fmfposition
\end{fmfgraph*}} =\frac{ f^{abc}}{e^2} 
\left(C\g_\m\right)_{\a\b} \d_{AB},
\qquad\parbox{20mm}{
\begin{fmfgraph*}(50,50)
\fmftop{v1}
\fmfbottom{v2,v3}
%\fmffreeze
\fmf{fermion,label=$p_1$}{v1,vc1}
\fmf{fermion,label=$p_2$,label.side=left}{vc1,v2}
\fmf{photon,label=$p_3$,label.side=right}{v3,vc1}
\fmflabel{$a$}{v1}
\fmflabel{$b$}{v2}
\fmflabel{$c,\m$}{v3}
%\fmffreeze
%\fmfposition
\end{fmfgraph*}} =\frac{ f^{abc}}{e^2} 
(p_1+p_2)_\m,
\]\\
\[
\parbox{20mm}{
\begin{fmfgraph*}(50,50)
\fmftop{v1}
\fmfbottom{v2,v3}
%\fmffreeze
\fmf{scalar,label=$p_1$}{v1,vc1}
\fmf{scalar,label=$p_2$,label.side=left}{vc1,v2}
\fmf{plain,label=$p_3$,label.side=right}{v3,vc1}
\fmflabel{$a,A,\a$}{v1}
\fmflabel{$b,B,\b$}{v2}
\fmflabel{$c$}{v3}
%\fmffreeze
%\fmfposition
\end{fmfgraph*}} =\frac{ f^{abc}}{e^2} 
C_{\a\b}\, \e_{AB}.\]\\

\ni 

\section{Calculational details} \label{4fermi}

In this appendix, we will compute the 4-fermion amplitudes of the YMCS
theory and the CSM theory. We first compute the 4-fermion YMCS
amplitudes, since the corresponding result in the CSM theory will then
follow straightforwardly.

In the YMCS theory, the basic building blocks for the four-fermion amplitudes are the gluon and scalar exchange, given by
\bsp
&{\cal A}(1,2,3,4)=
\bar v(p_1) \g^\m u(p_2)\,
\D_{\m\n}(p_1+p_2) \,
\bar v(p_3) \g^{\n}u(p_4),
\end{split}
\ee
where $\D_{\m\n}$ is the YMCS gauge field propagator
(see (\ref{props})), and
\bsp
&{\cal B}(1,2,3,4)=
\bar v(p_1) u(p_2) 
\,\frac{1}{(p_1+p_2)^2+m^2} \,
\bar v(p_3) u(p_4),
\end{split}
\ee
respectively. Defining the colour-ordered amplitudes $\la \phi_{{\cal A}_1}\phi_{{\cal A}_2} \phi_{{\cal A}_3}\phi_{{\cal
    A}_4}\ra$  of
completely general fields $\phi_{{\cal A}_i}$ as 
\bsp
\Bigl\la {\phi^{a_1}_{{\cal A}_1}}^\dag(p_1)\,& {\phi^{a_2}_{{\cal A}_2}}^\dag(p_2)\, 
{\phi^{a_3}_{{\cal A}_3}}^\dag(p_3)\, {\phi^{a_4}_{{\cal
      A}_4}}^\dag(p_4)\, \Bigr\ra\\
& = 2ie^2\, \left\la\phi_{{\cal A}_1}\phi_{{\cal A}_2} \phi_{{\cal A}_3}\phi_{{\cal
    A}_4} \right\ra\Tr[T^{a_1}T^{a_2}T^{a_3}T^{a_4}] + \ldots,
\end{split}
\ee
we find that
\bsp
\left\la\Psi_{A_1}\Psi_{A_2}\Psi_{A_3}\Psi_{A_4}\right\ra
=
& \d_{A_1A_2} \d_{A_3A_4} \Bigl( {\cal B}(4,1,2,3)+
 {\cal A}(1,2,3,4)
 \Bigr) \\
-& \d_{A_1A_3} \d_{A_2A_4} \Bigl(
{\cal B}(4,1,2,3) -  {\cal B}(1,2,3,4)\Bigr)\\
-& \d_{A_1A_4} \d_{A_2A_3}\Bigl( 
{\cal A}(4,1,2,3)
+  {\cal B}(1,2,3,4)\Bigr).
\end{split}
\ee
The expressions for the gluon and scalar exchange may be compactly
expressed as follows
\bsp\label{scrAB}
&{\cal B}(1,2,3,4) = \frac{\la 1 2\ra\la 3 4\ra}{(p_1+p_2)^2+m^2},\\
&{\cal A}(1,2,3,4) =  \frac{1}{(p_1+p_2)^2+m^2}\Biggl(
2\la  2 3\ra\la 4 1\ra-
\la  1 2\ra\la 3 4\ra
-2im\,\frac{\la 1 3\ra\la 14\ra}{\la1\bar 2\ra}\Biggr).
\end{split}
\ee

Because of the inherent $SO(2)$ symmetry enjoyed by the fermions, it
is useful to make the combinations
\be
\chi_\pm = \frac{1}{\sqrt{2}}\left(\Psi_1 \pm i\Psi_2\right),
\ee
which gives rise to the following amplitudes\footnote{We define
  $s=(p_1+p_2)^2$, $t=(p_1+p_3)^2$, $u=(p_1+p_4)^2$.}
\bsp
&\la\chi_+\chi_+\chi_-\chi_-\ra =\la\chi_-\chi_-\chi_+\chi_+\ra =
 -{\cal A}(4,1,2,3) -{\cal B}(4,1,2,3)\\
&\qquad\qquad\qquad\qquad\qquad\qquad\quad
=-\frac{2\la 3 4\ra}{u+m^2}\left[\la 1
  2\ra+im\frac{\la 4 2\ra}{\la 4 \bar 1\ra}\right],\\
&\la\chi_+\chi_-\chi_-\chi_+\ra =\la\chi_-\chi_+\chi_+\chi_-\ra = {\cal A}(1,2,3,4) + 
{\cal B}(1,2,3,4)\\
&\qquad\qquad\qquad\qquad\qquad\qquad\quad
=\frac{2\la4 1\ra}{s+m^2}\left[\la 2
  3\ra+im\frac{\la 1 3\ra}{\la 1 \bar 2\ra}\right],\\
&\la\chi_+\chi_-\chi_+\chi_-\ra =\la\chi_-\chi_+\chi_-\chi_+\ra =
{\cal A}(1,2,3,4) - {\cal B}(1,2,3,4)\\
&\qquad\qquad\qquad\qquad\qquad\qquad\quad
 - {\cal A}(4,1,2,3) + {\cal B}(4,1,2,3)\\
&
=\frac{2\la13\ra}{s+m^2}\left[\la 4
  2\ra-im\frac{\la14 \ra}{\la 1 \bar 2\ra}\right]
-\frac{2\la42\ra}{u+m^2}\left[\la 3
  1\ra-im\frac{\la43 \ra}{\la 4 \bar 1\ra}\right].
\end{split}
\ee

The calculation of the colour-ordered four-fermion amplitudes of the CSM
theory is similar to the one we carried out for the YMCS theory. In
fact the $\langle \chi_+\chi_+\chi_-\chi_-\rangle$ amplitude may be read-off from (\ref{scrAB}). There is no Yukawa
coupling in the CSM theory, thus the tree-level four-fermion amplitudes
are given only by the exchange of the gauge field. Thus we can take
${\cal B}$ to zero, and take the $m\to\infty$ limit in ${\cal A}$, in
order to single-out the pure CS term in the YMCS gauge field
propagator. This corresponds to keeping only the last term in ${\cal
  A}$, and replacing $(p_1+p_2)^2+m^2\to m^2$ in the factor outside
the rounded brackets. Multiplying by $e^2$ and noting that $e^2/m=\kappa$, where $\kappa=k/4\pi$, one then finds that the 4-fermion amplitude of the
CSM theory is given by
\begin{equation}
\langle \chi_+\chi_+\chi_-\chi_-\rangle_{CSM} = -2i\frac{\la 3 
4\ra\la 4  2\ra}
{\la 4\bar 1\ra},
\ee
where we absorbed $\kappa$ into the normalization of the fields.

%%%%%%%%%%%%%%%%%%%%%%%%%%%%%%%%%%%%%%%%%
\section{BCFW details}
\label{sec:fact}

We note that the BCFW shift has the following form on momenta:
\be
p_{\substack{i\\j}} \to \frac{1}{2}\left(p_i+p_j\right) \pm z^2 q \pm
z^{-2} \tilde q,
\ee
so that $p_i+p_j \to p_i+p_j$. We find that $q$ and $\tilde q$ may be
parameterized as follows
\bsp
&q = \frac{1}{4} \left( p_i - p_j + \frac{2}{\sqrt{s_{ij}}}
  p_i \wedge p_j \right), \\
&\tilde q = \frac{1}{4} \left( p_i - p_j - \frac{2}{\sqrt{s_{ij}}}
  p_i \wedge p_j \right) ,
\end{split}
\ee
where $(a\w b)^\m = \e^{\m\n\r} a_\n b_\r$, $s_{ij} = (p_i+p_j)^2$, and  
\bsp
&q\cdot (p_i + p_j) = \tilde q\cdot (p_i + p_j) =0 = q^2 = \tilde q^2,\\
&q+\tilde q = \frac{1}{2} \left(p_i - p_j\right).
\end{split}
\ee

We will be interested in the deformation of the remaining Mandelstam
invariants $s_{ik}$ and $s_{jk}$, where\footnote{The remaining
  momentum $p_4$ in the four particle process is equal to
  $-p_i-p_j-p_k$.} $i\neq j \neq k \in {1,2,3}$. We find
\bsp
s_{ik} \to \frac{1}{2}\left(s_{ik}+s_{jk}\right)
&+\frac{z^2}{2} \left(\frac{1}{2}(s_{ik} -
s_{jk})+\e_{ijk}\sqrt{-s_{ik}s_{jk}}\right)\\
&+\frac{z^{-2}}{2} \left(\frac{1}{2}(s_{ik} -
s_{jk})-\e_{ijk}\sqrt{-s_{ik}s_{jk}}\right),
\end{split}
\ee
\bsp
s_{jk} \to \frac{1}{2}\left(s_{ik}+s_{jk}\right)
&-\frac{z^2}{2} \left(\frac{1}{2}(s_{ik} -
s_{jk})+\e_{ijk}\sqrt{-s_{ik}s_{jk}}\right)\\
&-\frac{z^{-2}}{2} \left(\frac{1}{2}(s_{ik} -
s_{jk})-\e_{ijk}\sqrt{-s_{ik}s_{jk}}\right),
\end{split}
\ee
where we have used $p_i\cdot p_j\w p_k = \e_{ijk} \sqrt{-stu}$. We
will be looking for poles in the massive channels $s_{ik}+m^2$ and
$s_{jk} + m^2$; these correspond to the following equations
\bsp
\frac{z^4}{2} \left( s_-+\sqrt{s_-^2-s_+^2}\right)
\pm z^2(s_++1)+\frac{1}{2}\left( s_--\sqrt{s_-^2-s_+^2}\right) = 0,
\end{split}
\ee
where the upper sign corresponds to $s_{ik}$ and the lower to
$s_{jk}$, and where
\be
s_\pm = \frac{1}{2m^2}\left(s_{ik} \pm s_{jk}\right).
\ee
The roots of these two equations are
\be\label{roots}
\{z_1^2,z_2^2\} = \left\{\mp\frac{1}{2}\frac{\left(\sqrt{2s_++1}\mp 1\right)^2}
{s_-+\sqrt{s_-^2-s_+^2}},\,
\mp\frac{1}{2}\frac{\left(\sqrt{2s_++1}\pm 1\right)^2}
{s_-+\sqrt{s_-^2-s_+^2}}
\right\}.
\ee

\end{fmffile}
%%%%%%%%%%%%%%%%%%%%%%%%%%%%%%%%%%%%%%%%%%%%%%%%%%%%%%%%%%%%%%%%%%%%%

\end{document}